\UseRawInputEncoding
\documentclass[5p]{elsarticle}

\journal{Elsevier}

\usepackage{subfigure}
\usepackage{fancyhdr}
\usepackage{amsmath}
\usepackage{multirow}
\usepackage{amssymb }
\usepackage{color}
\usepackage{graphics} 
\usepackage{graphicx}
\usepackage{amsmath}
\newtheorem{theorem}{\textbf{Theorem}}
\newtheorem{lemma}{\textbf{Lemma}}
\newtheorem{example}{\textbf{Example}}
\newtheorem{corollary}{\textbf{Corollary}}
\newtheorem{remark}{\textbf{Remark}}
\newtheorem{definition}{\textbf{Definition}}

\newtheorem{proposition}{\textbf{Proposition}}

\usepackage{url}

\newenvironment{proof}{{{\bf Proof:}}}{\hfill $\square$\par}
\usepackage{multirow} 
\usepackage{xcolor}
\usepackage{multirow}
\usepackage{setspace}
\usepackage{url}
\usepackage{subfigure}
\usepackage{fancyhdr}
\usepackage{amsmath}
\usepackage{multirow}
\usepackage{amssymb }
\usepackage{color}
\usepackage{graphics} 
\usepackage{graphicx}
\usepackage{algorithm} 
\usepackage{algorithmic} 
\usepackage{subeqnarray}
\usepackage{cases}
\usepackage{threeparttable}

\begin{document}
	\begin{frontmatter}
		\title{On the Reachability and Controllability of Temporal \\~  Continuous-Time Linear Networks: A Generic Analysis}
		
		
		
		\author[author1]{Yuan Zhang}
		
		\author[author1]{Yuanqing Xia}
		
		\author[author2]{Long Wang}
		
		\address[author1]{School of Automation, Beijing Institute of Technology, Beijing, China}
		
		\address[author2]{Center for Systems and Control, College of Engineering, Peking University, Beijing, China}
		
		\address{Email: $\emph{zhangyuan14@bit.edu.cn, xia\_yuanqing@bit.edu.cn, longwang@pku.edu.cn}$} 
		
		\fntext[myfootnote]{This work was supported in part by the
			National Natural Science Foundation of China under Grant 62003042. } 

		\begin{abstract}Temporal networks are a class of time-varying networks, which change their topology according to a given time-ordered sequence of static networks (known as subsystems). This paper investigates the reachability and controllability of temporal {\emph{continuous-time}} linear networks from a generic viewpoint, where only the zero-nonzero patterns of subsystem matrices are known.
			We demonstrate that the reachability and controllability on a single temporal sequence are generic properties with respect to the parameters of subsystem matrices and the {\emph{time durations}} of subsystems. We then give explicit expressions for the minimal subspace that contains the reachable set across all possible temporal sequences (called overall reachable set). It is found that verifying the structural reachability/controllability and structural overall reachability are at least as hard as the structural target controllability verification problem of a single system, implying that finding verifiable conditions for them is hard.  Graph-theoretic lower and upper bounds are provided for the generic dimensions of the reachable subspace on a single temporal sequence and of the minimal subspace that contains the overall reachable set. These bounds extend classical concepts in structured system theory, including the dynamic graph and the cactus, to temporal networks, and can be efficiently calculated using graph-theoretic algorithms. Finally, applications of the results to the structural controllability of switched linear systems are discussed.
			
			%
			%
			
		\end{abstract}
		
	\end{frontmatter}
	
	%

	\section{Introduction} \label{intro-sec}
	Temporal networks are a class of time-varying networks that change their typologies (including changes on node connectivity and edge weights) according to a fixed sequence of static networks \cite{posfai2014structural,li2017fundamental}. Compared to the static network model, topology temporality has been recognized to ubiquitously exist in dynamic processes taking place on networks \cite{holme2005network,barabasi2005origin}. For example, in social networks, contacts among individuals are often intermittent and recurrent, rather than persistent \cite{kossinets2008structure}. On the Internet, data transmissions between agents are typically bursty and fluctuate over time \cite{barabasi2005origin}. In wireless communication networks, some resource-aware control strategies such as the event-trigger communication \cite{wei2022distributed} may also lead to link temporality.  Other examples can be found in the brain network, transportation networks \cite{holme2005network}, and so on.  It has been found that, although link temporality may disrupt network reachability, slowing down synchronization and consensus of innovative information \cite{holme2005network}, it does have advantages over the static networks, including reaching controllability faster and demanding less control energy \cite{li2017fundamental}.
	
	In network science, controllability (or reachability) is the ability to steer the node states arbitrarily in the whole space by choosing suitable inputs. Network controllability has attracted considerable attention during the last decade \cite{Y.Y.2011Controllability,Ramos2022AnOO}. Significant achievements have been reached on understanding how network topology (such as degree distributions and connectivity) and nodal dynamics influence system controllability \cite{Y.Y.2011Controllability,zhang2019structural}, as well as the related optimal input/output selection and topology design problems \cite{S.Pe2016A, zhang2019minimal}; see \cite{Ramos2022AnOO} for a survey. Yet, most of these works focus on static networks whose topology is invariant over time. Understanding the controllability of temporal networks is more challenging but often a prerequisite for analyzing more complicated processes on networks, including consensus and synchronization \cite{li2017fundamental,zhang2021unified}.

	On the other hand, from the system theory,  temporal networks can be seen as switched systems when the switching sequence is predetermined.
	In the context of hybrid systems and switched systems, controllability and reachability have been extensively studied \cite{conner1987structure,ezzine1989controllability,sun2001reachability,Z.S2002Controllability,xie2003controllability}. When the switching law can be designed arbitrarily, it is found the reachable and controllable set of switched continuous-time systems are subspaces of the total space, and the complete characterizations were given in \cite{Z.S2002Controllability}. For discrete-time systems, the reachable and controllable set are not necessarily subspaces \cite{stanford1980controllability}. Geometric characterizations for these sets to be the total space were presented in \cite{ge2001reachability}. Some necessary/sufficient conditions for the controllability of periodic hybrid systems were given in \cite{ezzine1989controllability}. These results rely on algebraic computations involving the precise values of system matrices. However, the exact system parameters may be hard to know in practice due to
	modeling errors or parameter uncertainties \cite{Y.Y.2011Controllability}. In this case, when only the sparsity
	patterns of system matrices are available, \cite{LiuStructural} introduced the notion {\emph{structural controllability}} for {\emph{switched systems}}, an alternative notion of controllability in the generic sense initiated by \cite{Lin_1974}. Some colored-graph based criteria were given in \cite{LiuStructural}. The optimal input selection problems for structural controllability of switched systems were investigated in \cite{pequito2017structural,zhang2022constrained}.
	
	
	Structural controllability of temporal {\emph{discrete-time}} networks has been well-understood \cite{poljak1990generic,posfai2014structural,zhang2021higher}. Particularly, when each subsystem appears only once in the discrete-time iterations, the so-called {\emph{dynamic graph}} can completely characterize the structural controllability of temporal networks and the associated controllable/reachable subspaces. However, systems that evolve in continuous time naturally arise in the real world. Although an attempt has been made in \cite{hou2016structural}, the full characterization of structural controllability of temporal {\emph{continuous-time} networks is far from reached.  
		This is due to two factors: the time duration of each subsystem in continuous-time networks is often not fixed, and the controllable/reachable set over all time durations is an uncountable union of subspaces, which is not necessarily a subspace. 
		
		In this paper, we investigate the structural reachability and controllability of temporal continuous-time networks, addressing these challenges by providing refined definitions of structural reachability and structural overall reachability that avoid the dependence on subsystem time durations. Our main contributions are as follows. Firstly, we show that the dimension of reachable and controllable subspaces on a single temporal sequence is a generic property with respect to both the parameters of subsystem matrices and the time duration of each subsystem. We also give explicit geometric and algebraic expressions for the minimal subspace (denoted by $\bar {\bf{\Omega}}$) containing the reachable/controllable set over all possible temporal sequences. Secondly, we demonstrate that verifying structural reachability (controllability) and structural overall reachability (controllability) is at least as hard as the structural target controllability verification problem (STCP), for which there is no known verifiable criterion. Thirdly, we develop graph-theoretic upper and lower bounds for the generic dimensions of reachable subspace on a single temporal sequence and the minimal subspace that contains the reachable/controllable set over all possible temporal sequences. These bounds can be computed efficiently using graph-theoretic algorithms, which are attractive for large-scale systems.  Additionally, in obtaining them, several classical notions in structured (or structural) system theory, namely, the dynamic graph and the cactus, are extended to temporal networks. To the best of our knowledge, this is the first time that these tools are applied to temporal continuous-time networks. Finally, we show that these bounds can be applied to the controllability realization problem and provide bounds for the generic dimension of the controllable subspace of switched systems. Compared to the switched systems \cite{Z.S2002Controllability,xie2003controllability,LiuStructural}, our work also includes some interesting findings, such as: (1) the overall reachability of a temporal network does not necessarily imply the existence of a temporal sequence on which the reachability is achieved; (2) the overall reachability of a temporal network is not necessarily generic w.r.t. the subsystem parameters.  Our work contributes to a graph-theoretic approach to understanding how subsystem structure and time duration influence the controllability and reachability of temporal continuous-time networks, with potential applications in the design of switched systems.

		The remaining sections are organized as follows. Section \ref{sec-2} presents the model description and relevant definitions. In Section \ref{sec-3}, we provide an elementary analysis of the reachable set and justify the involved generic properties. Section \ref{sec-4} proves that verifying structural reachability and related problems are at least as difficult as the STCP. Sections \ref{sec-5} and \ref{sec-6} respectively provide graph-theoretic upper and lower bounds for the generic dimensions of the reachable subspace and the subspace $\bar{\bf{\Omega}}$. In Section \ref{sec-application}, we briefly discuss the applications of the previous results to structured switched systems. The final section concludes this paper.
		
		
		{\bf Notations:}	$I_n$ denotes the $n\times n$ identify matrix.  For a matrix $M$, $M_{ij}$ (or sometimes $M(i,j)$) denotes the $(i,j)$th entry of $M$. The notation ${\bf row}\{X_i|_{i=1}^l\}$ represents a  composite matrix whose column blocks are $X_1,\ldots,X_l$.  For a matrix $M$, ${\rm span} M$ is the space spanned by the columns of $M$. For a subspace ${\bf \Omega}\subseteq {\mathbb R}^n$, ${\rm dim} {\bf \Omega}$ is its dimension. ${\mathbb R}^+$ denotes the set of positive real values.   
		
		\section{Model description and definitions} \label{sec-2}
		A temporal sequence is a set of successive time intervals $\{[t_{i-1}, t_i)\}_{i=1}^N$, where $t_i >t_{i-1}$ for $i=1,\cdots, N$, $N<\infty$.\footnote{It is worth noting that all results of this paper are still valid even when we allow $t_i\ge t_{i-1}$.} 
		Temporal systems are a class of time-varying systems, in which there is a temporal sequence $\{[t_{i-1}, t_i)\}_{i=1}^N$ such that in each time interval $[t_{i-1},t_i)$, the system is governed by the following linear time-invariant dynamics: 
		\begin{equation} \label{temporal-plant}
		\dot x(t)=A_ix(t)+B_iu_i(t),
		\end{equation}
		where $x(t)\in {\mathbb{R}}^n$ is the state, $u_i(t)\in {\mathbb R}^{m_i}$ is the piecewise continuous input, and $n$ and $m_i$ are respectively the number of state variables and inputs for $t\in [t_{i-1}, t_i)$. In other words, system (\ref{temporal-plant}), denoted also by $(A_i,B_i)$, is activated as the system realization in the $i$th `snapshot' $[t_{i-1},t_i)$. Here, $(A_i, B_i)$ is called a subsystem, $i=1,...,N$. In some literature, the $n$ state variables represent $n$ nodes of a network whose topology changes over time (called temporal networks \cite{nicosia2013graph}). More precisely, $A_i$ is the weighted adjacency matrix of the network for the snapshot $[t_{i-1}, t_i)$ so that $A_{i,jk}\ne 0$ means there is a directed edge from node $k$ to node $j$, and $B_i$ is the corresponding input matrix such that $B_{i,jk}\ne 0$ means node $j$ is directly actuated by the $k$th input,  for $t\in [t_{i-1},t_i)$. Then, system (\ref{temporal-plant}) can describe the dynamics of a {\emph{temporal network}} \cite{posfai2014structural,li2017fundamental}.
		
		If the temporal sequences can be designed arbitrarily without a specified order, that is, the switching between any two $(A_i, B_i)$ and $(A_j, B_j)$ is allowable ($i,j=1,...,N$), we obtain
		the switched systems \cite{conner1987structure}. More precisely, a switched system is described by the following equation \cite{Z.S2002Controllability}
		\begin{equation} \label{plant-switched}
		\dot x(t)= A_{\sigma(t)} x(t) + B_{\sigma(t)} u_{\sigma(t)}(t),
		\end{equation}
		where $x(t)\in {\mathbb R}^n$ is the state, $\sigma(t): [0,\infty) \to \{1,...,N\}$ is the switching signal, $u_{\sigma(t)}(t)\in {\mathbb R}^{m_i}$ is the piecewise continuous input,
		and $\sigma(t)=i$ implies the subsystem $(A_i, B_i)$ is activated as the system realization at time instant $t$. A switching path is a sequence of subsystem indices in a switching signal.  We may use the matrix set $(A_i,B_i)|_{i=1}^N$ to denote system (\ref{temporal-plant}) and system (\ref{plant-switched}).  As can be seen,  temporal systems (\ref{temporal-plant}) can be regarded as switched systems (\ref{plant-switched}) with fixed switching signals.
		
		Given a temporal sequence $\Pi\doteq \{[t_{i-1}, t_i)\}_{i=1}^N$, let $h_i=t_i-t_{i-1}$ be the time duration of the $i$th subsystem, $i=1,...,N$. We may also use $\{h_i\}_{i=1}^N$ (sometimes $\{h_i\}$) to indicate the temporal sequence $\Pi$.   
		We introduce the notion of reachability (controllability) for system (\ref{temporal-plant}) as follows.
		
		\begin{definition}[Reachable state on $\Pi$]\cite{li2017fundamental} \label{reachable-point-on-sequence}
			For system (\ref{plant-switched}), a state $x\in  {\mathbb R}^n$ is said to be reachable (resp. controllable) on a temporal sequence $\Pi$ if there exist piecewise continuous functions: $u_i(t): [t_{i-1},t_i)\to {\mathbb R}^{m_i}$, $i=1,...,N$, such that $x(t_0)=0$ and $x(t_N)=x$ ($x(t_0)=x$ and $x(t_N)=0$).
		\end{definition}
		
		\begin{definition}[Reachable set on $\Pi$]\cite{li2017fundamental}\label{reachable-set-on-sequence} The reachable set (resp. controllable set) of system (\ref{temporal-plant}) on the temporal sequence ${\Pi}$ is the set of states which are reachable (controllable) on $\Pi$. If this reachable (controllable) set is ${\mathbb R}^n$, system (\ref{temporal-plant}) is said to be reachable (controllable) on~$\Pi$.
		\end{definition}
		
		Note that the above definitions depend on a specific temporal sequence $\Pi$. When $\Pi$ is not specified, we introduce the following definitions.
		
		\begin{definition} \label{def-reachable-point-temporal} For system (\ref{temporal-plant}), state $x\in  {\mathbb R}^n$ is reachable (resp. controllable), if there exist a temporal sequence $\{[t_{i-1}, t_i)\}_{i=1}^N$ and inputs $u_i(t): [t_{i-1},t_i)\to {\mathbb R}^{m_i}$, $i=1,...,N$, $t_N<\infty$, such that $x(t_0)=0$, and $x(t_N)=x$ ($x(t_0)=x$ and $x(t_N)=0$).
		\end{definition}
		
		\begin{definition}[Overall reachable set] \label{reachable-set1} The overall reachable set (resp. overall controllable set) of system (\ref{temporal-plant}) is the set of states which are reachable (controllable). If the overall reachable (controllable) set is ${\mathbb R}^n$, system (\ref{temporal-plant}) is said to be overall reachable (overall controllable).
		\end{definition}
		
		
		Roughly speaking,  the overall reachable set is the union of reachable sets on all possible temporal sequences. Hereafter, for ease of descriptions, {\emph{by referring to `reachable (set)' we mean `reachable (set)' on some temporal sequence $\Pi$, if not specified.}} Since we shall discuss the relations of our results with those of switched systems, for self-consistency, the related definitions for switched systems are introduced as follows.
		
		\begin{definition} \cite{Z.S2002Controllability} \label{def-reachable-point-switched} For system (\ref{plant-switched}), state $x\in  {\mathbb R}^n$ is reachable (resp. controllable), if there exist a finite time interval $[t_0,t_f)$ ($t_f>t_0$), a switching signal $\sigma(t): [t_0, t_f)\to \{1,...,N\}$, and piecewise continuous input functions $u_{\sigma(t)}(t)$ $\in {\mathbb R}^{m_{\sigma(t)}}$ for $t\in [t_0,t_f)$, such that $x(t_0)=0$ and $x(t_f)=x$ (resp. $x(t_0)=x$ and $x(t_f)=0$). The reachable (controllable) set of system (\ref{plant-switched}) is the set of states which are reachable (controllable). If the reachable (controllable) set is ${\mathbb R}^n$, system (\ref{plant-switched}) is said to be reachable (controllable).
		\end{definition}
		
		Notably, for switched systems, an equivalent definition of the reachable set in Definition \ref{def-reachable-point-switched} is the maximum dimensional subspace of the total space that can be steered to from the zero initial state on a {\emph{single switching signal}} (see \cite{Z.S2002Controllability,xie2003controllability} for detail). However, for temporal networks, as we shall show in Section \ref{sec-genericity}, things are different. This is perhaps because, the switching path $(1,2,...,N)$ is fixed but only the duration $\{h_i\}_{i=1}^N$ can be chosen thereof, while for switched systems, both are free to be chosen. It is also proven in \cite{Z.S2002Controllability}, for switched systems, the reachable set and the controllable set always coincide with each other.
		

		
		In practical scenarios, the exact numerical values of $(A_i,B_i)|_{i=1}^N$ (and even the duration $\{h_i\}_{i=1}^N$) may be hard to know. By contrast, their sparsity patterns, i.e., the zero and non-zero patterns of matrices $(A_i,B_i)|_{i=1}^N$, may be easier to obtain.  In this case, it is preferable to characterize the generic properties of these reachable sets, that is, properties that hold with probability one, i.e., for almost all\footnote{A property is said to hold for {\emph{almost all}} of the considered parameters if the set for which this property does not hold has Lebesgue measure zero.} indeterminate parameters in $(A_i,B_i)|_{i=1}^N$ (and the duration $\{h_i\}_{i=1}^N$), which is the primary goal of this paper.
		
		\section{Elementary analysis and genericity justification} \label{sec-3}
		In this section, we characterize the reachable (controllable) sets of numerically specified temporal systems and justify the involved generic properties. We also highlight the difference between (overall) reachability of a temporal network and that of a switched system. 
		
		\subsection{Preliminary derivations} \label{sec-preliminary}
		Recall $\{h_i\}_{i=1}^N$ is the duration of the temporal sequence $\{[t_{i-1}, t_i)\}_{i=1}^N$. Starting from the initial state $x(t_0)$, the state $x(t_N)$ at time instant $t_N$ of system (\ref{temporal-plant}) is \cite{Z.S2002Controllability}
		\begin{equation}\label{reachable-point}
		\begin{array}{c}
		x(t_N)= \prod \limits_{i=1}^{N}e^{A_ih_i}x(t_0)+\prod \limits_{i=2}^{N}e^{A_ih_i}\int_{t_0}^{t_1}e^{A_1(t_1-\tau)}B_1u_1(\tau)d\tau \\
		+ \cdots+ \int_{t_{N-1}}^{t_N}e^{A_N(t_{N}-\tau)}B_Nu_N(\tau)d\tau.
		\end{array}	
		\end{equation}
		According to \citep[Lemma 2.10]{antsaklis1997linear}, for any $A\in {\mathbb R}^{n\times n}$, $B\in {\mathbb{R}}^{n\times m}$ and $t>t_0\ge 0$,
		\begin{equation}\label{reachable-formula}\{x:x=\int_{t_0}^{t}e^{A(t-\tau)}Bu(\tau)d\tau, u(\tau)\in {\mathbb{R}}^m\}=\left \langle A|B\right \rangle,\end{equation}
		where $\left \langle A|B\right \rangle=\sum_{i=0}^{n-1}A^i{\rm Im} B$, with ${\rm Im} B$ the subspace spanned by the columns of $B$. By substituting $x(t_0)=0$ to (\ref{reachable-point}), the reachable set of system (\ref{temporal-plant}) on $\{h_i\}_{i=1}^N$ is~\cite{Z.S2002Controllability}
		\begin{equation} \label{reachable-set-time-dependent}
		{\mathbf \Omega}_{\{h_i\}}=\left \langle A_N|B_N\right \rangle+\sum_{j=2}^{N}\prod_{i=j}^{N}e^{A_ih_i}\left \langle A_{j-1}|B_{j-1}\right \rangle,
		\end{equation}
		where $\prod_{i=j}^Ne^{A_ih_i}=e^{A_Nh_N}e^{A_{N-1}h_{N-1}}\cdots e^{A_jh_j}$.
		The overall reachable set ${\mathbf \Omega}$ of system (\ref{temporal-plant}) is the union of reachable sets on all possible temporal sequences, thus expressed~as
		\begin{equation} \label{reachable-set}
		{\mathbf \Omega}=\bigcup\limits_{h_1,...,h_N>0} \left( \left \langle A_N|B_N\right \rangle+\sum_{j=2}^{N}\prod_{i=j}^{N}e^{A_ih_i}\left \langle A_{j-1}|B_{j-1}\right \rangle \right).
		\end{equation}
		Obviously, ${\mathbf{\Omega}}_{\{h_i\}}$ is a subspace of ${\mathbb{R}}^n$, while ${\mathbf{\Omega}}$ is necessarily not since it is an uncountable union of subspaces of ${\mathbb{R}}^n$.
		The following typical example illustrates this.
		\begin{example} \label{not-subspace-exp}
			Consider a temporal network with
			
			{\small $$A_1=\left[\begin{array}{ccc}0 & 0 & 0 \\
				1 & 0 & 0\\
				0 & 0 & 0
				\end{array}\right],B_1=\left[\begin{array}{c}1\\
				0\\
				0
				\end{array}\right],$$ $$A_2=\left[\begin{array}{ccc}0 & 0 & 0 \\
				0 & 0 & 0\\
				0 & 1 & 0
				\end{array}\right], B_2=\left[\begin{array}{c}0\\
				0\\
				0
				\end{array}\right].$$}Simple calculations show that
			$${\mathbf{\Omega}}_{\{h_i\}}=\left\{\left[\begin{array}{c}x_1\\
			x_2\\
			h_2x_2
			\end{array}\right]: x_1,x_2\in{\mathbb R}\right\},$$
			$${\mathbf{\Omega}}=\left\{\left[\begin{array}{c}x_1\\
			x_2\\
			h_2x_2
			\end{array}\right]: x_1,x_2\in{\mathbb R},h_2>0\right\}.$$Note ${\mathbf{\Omega}}_{\{h_i\}}$ is a subspace of ${\mathbb{R}}^3$ with dimension $2$ for any $h_1,h_2>0$. By contrast,  ${\mathbf{\Omega}}$ is neither a subspace nor a countable union of subspaces (the second coordinate and the third one of every vector in ${\mathbf{\Omega}}$ have the same sign).
		\end{example}
		
		Similar to the reachable set (\ref{reachable-set}), the controllable set of system (\ref{temporal-plant}) on $\{h_i\}_{i=1}^N$ and the overall controllable set are given by \cite{Z.S2002Controllability,xie2003controllability}
		\begin{equation} \label{controllable-set}
		\begin{aligned}
		{\mathbf \Theta}_{\{h_i\}}&= \left \langle A_1|B_1\right \rangle+\sum_{j=2}^{N}\prod_{i=1}^{j-1}e^{-A_ih_i}\left \langle A_{j}|B_{j}\right \rangle \\
		{\mathbf \Theta}&=\bigcup\nolimits_{h_1,...,h_N>0} {\mathbf \Theta}_{\{h_i\}},
		\end{aligned}
		\end{equation}where $\prod_{i=1}^{j-1}e^{-A_ih_i}=e^{-A_1h_1}\cdots e^{-A_{j-1}h_{j-1}}$. From (\ref{controllable-set}), we know:
		\begin{itemize}
			\item[(1)] Given $\{h_i\}_{i=1}^N$, ${\mathbf{\Omega}}_{\{h_i\}}$ does not necessarily equal ${\mathbf \Theta}_{\{h_i\}}$;
			\item[(2)] ${\mathbf{\Omega}}$ does not necessarily equal ${\mathbf \Theta}$;
			\item[(3)] For any $\{h_i\}_{i=1}^N$, ${\rm dim}{\mathbf{\Omega}}_{\{h_i\}}={\rm dim}{\mathbf \Theta}_{\{h_i\}}$. \end{itemize}
		To see (1) and (2), consider the temporal network in Example \ref{not-subspace-exp}. It turns out that ${\bf \Omega}_{\{h_i\}}=e^{A_2h_2}\langle A_1|B_1\rangle\ne \langle A_1|B_1\rangle={\mathbf \Theta}_{\{h_i\}}$, for any $h_2>0$. Moreover,  ${\mathbf \Theta}=\langle A_1|B_1\rangle$, which is a subspace of ${\mathbb{R}}^3$, while ${\mathbf{\Omega}}$ is not a subspace.
		To see (3), by virtue of ${\rm span}e^{A_ih_i}\langle A_i|B_i \rangle={\rm span}\langle A_i|B_i \rangle$ for any $h_i>0$ from (\ref{reachable-formula}), we have $e^{A_{N}h_N}e^{A_{N-1}h_{N-1}}\cdots e^{A_1h_1}{\mathbf \Theta}_{\{h_i\}}={\bf \Omega}_{\{h_i\}}$.  Note that (3) has also been observed in \cite{li2017fundamental}.

		Two questions arise naturally: (1) how the dimension of ${\mathbf{\Omega}}_{\{h_i\}}$ depends on $\{h_i\}$ and how to achieve the maximum dimension, and (2) how to characterize the overall reachable set ${\mathbf{\Omega}}$.
		For the first question, we demonstrate that for almost all $h_1,...,h_N\in {\mathbb R}$, ${\mathbf{\Omega}}_{\{h_i\}}$ will have the same dimension, which is exactly the maximum dimension ${\mathbf{\Omega}}_{\{h_i\}}$ can take (Section \ref{sec-genericity}). For the second question, since ${\mathbf{\Omega}}$ is not necessarily a subspace,
		a common way is to find the minimal subspace that contains ${\mathbf{\Omega}}$, that is, a subspace $\bar {\mathbf{\Omega}}\subseteq {\mathbb{R}}^n$ that contains every element of ${\mathbf{\Omega}}$, and any proper subspace $\bar {\mathbf{\Omega}}'\subsetneqq \bar {\mathbf{\Omega}}$ cannot posses this property (Section \ref{sec-minimal-subspace}), which plays a role similar to the convex hull for a non-convex set \cite{antsaklis1997linear}. The generic properties related to ${\mathbf{\Omega}}_{\{h_i\}}$ and ${\mathbf{\Omega}}$ are justified alongside.  
		
		\subsection{Genericity of  ${\mathbf{\Omega}}_{\{h_i\}}$}		\label{sec-genericity}
		Before presenting the generic properties of ${\mathbf{\Omega}}_{\{h_i\}}$, the following definitions about structured (or structural) matrices are introduced.   
		
		\begin{definition}\cite{generic}
			A {\bf structured matrix} is a matrix whose entries are either fixed zero or not fixed to be zero (for simplicity, the latter is called a nonzero entry). Two structured matrices are said to be {\bf structurally equivalent}, if their zero-nonzero patterns are the same (but the nonzero entries in different matrices can take independent values).
			A  {\bf realization} of a structured matrix is a matrix obtained by assigning some particular values to the nonzero entries and preserving the zero entries.
		\end{definition}
		
		Let $C_i$ be the controllability matrix of the subsystem $(A_i,B_i)$, i.e., $C_i=[B_i,A_iB_i,...,A_i^{n-1}B_i]$, $i=1,...,N$. Define the reachability matrix as
		\begin{equation}\label{reachability-matrix}{\cal C}\!=\![C_N, e^{A_Nh_N}C_{N-1},\cdots, e^{A_Nh_N}\cdots e^{A_2h_2}C_1].\end{equation}
		Here, for notation simplicity, we omit the dependence of ${\cal C}$ on $\{h_i\}$. Obviously, the dimension of ${\bf \Omega}_{\{h_i\}}$ equals ${\rm rank} {\cal C}$.
		
		At first, we observe that given $(A_i,B_i)|_{i=1}^N$, the dimension of ${\bf \Omega}_{\{h_i\}}$ is generic w.r.t. $h_1,...,h_N$.\footnote{We say a property is generic w.r.t. a set of parameters if this property holds for almost all (i.e., all except a set of measure zero) values that the parameters can take. } 
		\begin{proposition} \label{generic-dimension}
			Given $(A_i,B_i)|_{i=1}^N$, the dimension of ${\bf \Omega}_{\{h_i\}}$ is the same for almost all $h_1,...,h_N\in {\mathbb R}$,\footnote{That is, $(h_1,...,h_N)$ can take values from ${\mathbb R}^N$ except for a set of measure zero. Note that we can restrict $h_1,...,h_N$ to be in any continuous interval, including {\emph{negative}} values.} and equals the maximum dimension it can achieve from all possible values of $h_1,...,h_N\in {\mathbb R}$.
		\end{proposition}
		\begin{proof}
			Write ${\cal C}$ as a function  $C(h_1,...,h_N)$ of $h_1,...,h_N$. Suppose $C(h_1,...,h_N)$ takes the maximum rank $r$ when $(h_1,...,h_N)$ takes values $(h_1^0,...,h_N^0)$. Then, there is a non-singular $r\times r$ sub-matrix with rank $r$ in $C(h^0_1,...,h^0_n)$. Denote the corresponding sub-matrix of $C(h_1,...,h_N)$ as $M(h_1,...,h_N)$, and its determinant as $d(h_1,...,h_N)$. Since
			every entry of $M(h_1,...,h_N)$ is an analytic function of $h_1,...,h_N$, $d(h_1,...,h_N)$ is also an analytic function of $h_1,...,h_N$.
			As $d(h_1^0,...,h_N^0)\ne 0$, $d(h_1,...,h_N)$ is not identically zero. According to \citep[Theorem 62]{kaplan1966introduction}, the zeros of $d(h_1,...,h_N)$ form a set of measure zero. Hence,
			for almost all $h_1,...,h_N\in {\mathbb R}$, $d(h_1,...,h_N)$ is not zero, which means $C(h_1,...,h_N)$ takes the rank $r$. This leads to that for almost all $h_1,...,h_N\in {\mathbb R}$, ${\bf \Omega}_{\{h_i\}}$ has the same dimension, which is the maximum dimension it takes as a function of $h_1,...,h_N$.
		\end{proof}
		
		Suppose only the sparsity patterns of $(A_i,B_i)|_{i=1}^N$ are known (that is, $(A_i,B_i)$ are regarded as structured matrices), and nonzero entries of $(A_i,B_i)|_{i=1}^N$ can take values independently (including zero values). Assume that there are $d$ nonzero entries in $(A_i,B_i)|_{i=1}^N$, and let $\phi$ be a $d$-dimensional parameter vector for those nonzero entries. For each value of $\phi\in {\mathbb R}^d$, we get a realization of $(A_i,B_i)|_{i=1}^N$.  We observe that, the dimension of  ${\bf \Omega}_{\{h_i\}}$ is generic w.r.t both $\phi$ and $h_1,...,h_N$.
		\begin{proposition} \label{generic-dimension-2}
			Given the sparsity patterns of $(A_i,B_i)|_{i=1}^N$, the dimension of ${\bf \Omega}_{\{h_i\}}$ is the same for almost all $h_1,...,h_N\in {\mathbb R}$ and $\phi\in {\mathbb R}^d$, which takes the maximum dimension obtained from all values of $h_1,...,h_N\in {\mathbb R}$ and $\phi\in {\mathbb R}^d$.
		\end{proposition}
		
		\begin{proof} The proof follows a similar manner to that of Proposition \ref{generic-dimension}. 
			Write ${\cal C}$ as a function of $h_1,...,h_N$ and $\phi$, reading $C(h_1,...,h_N,\phi)$. Suppose $C(h_1,...,h_N,\phi)$ takes the maximum rank $r$ when $(h_1,...,h_N,\phi)$ take values \\~ $(h_1^0,...,h_N^0,\phi^0)$, with $\phi^0\in {\mathbb{R}}^d$. Then, there is a non-singular $r\times r$ sub-matrix in $C(h^0_1,...,h^0_n,\phi^0)$. Denote the associated sub-matrix of $C(h_1,...,h_N,\phi)$ as $M(h_1,...,h_N,\phi)$, and its determinant as $d(h_1,...,h_N,\phi)$. As every entry of $M(h_1,...,h_N,\phi)$ is an analytic function of variables $h_1,...,h_N$, $\phi$, $d(h_1,...,h_N,\phi)$ is also an analytic function of $h_1,...,h_N,\phi$. Since $d(h_1^0,...,h_N^0,\phi^0)\ne 0$, by \citep[Theorem 62]{kaplan1966introduction}, the zeros of $d(h_1,...,h_N,\phi)$ form a set of measure zero in ${\mathbb R}^{N+d}$. Therefore, for almost all $h_1,...,h_N\in {\mathbb R}$ and $\phi\in {\mathbb R}^d$, $d(h_1,...,h_N,\phi)\ne 0$, indicating that $C(h_1,...,h_N,\phi)$ takes the maximum rank $r$. 
		\end{proof}
		
		Following Proposition \ref{generic-dimension-2}, we call the maximum dimension that ${\bf \Omega}_{\{h_i\}}$ could achieve from all values of $h_1,...,h_N$ and $\phi$ the {\emph{generic dimension}} of the reachable subspace ${\bf \Omega}_{\{h_i\}}$, denoted by ${\rm gdim} {\bf \Omega}_{\{h_i\}}$. It is clear ${\rm gdim} {\bf \Omega}_{\{h_i\}}$ equals the dimension of ${\bf \Omega}_{\{h_i\}}$ for almost all $h_1,...,h_N\in {\mathbb R}$ and
		$\phi\in {\mathbb R}^d$. Similarly, the maximum rank that ${\cal C}$ could take as a function of $h_1,...,h_N$ and $\phi$ is called the {\emph{generic rank}} of $\cal C$, given by ${\rm grank}{\cal C}$, which is the rank it takes for almost all values of $h_1,...,h_N$ and $\phi$. Obviously, ${\rm gdim} {\bf \Omega}_{\{h_i\}}={\rm grank}{\cal C}$, and their value depends on {\emph{combinatorial}} properties of sparsity patterns of $(A_i,B_i)|_{i=1}^N$.		
		
		With the generic property identified above, the notion structural reachability/controllability of the temporal network (\ref{temporal-plant}) is defined as follows.
		
		\begin{definition}[Structural reachability] \label{def-structural}
			Given the sparsity patterns of $(A_i,B_i)|_{i=1}^N$, we say system (\ref{temporal-plant}) is structurally reachable (resp. structurally controllable), if there is a realization of $(A_i,B_i)|_{i=1}^N$ that is reachable (resp. controllable) on some temporal sequence.
		\end{definition}
		
		It turns out that if $(A_i,B_i)|_{i=1}^N$ is structurally reachable, then almost all realizations of $(A_i,B_i)|_{i=1}^N$ are reachable on almost {\emph{any}} temporal sequence.
		Since for any $\{h_i\}_{i=1}^N$, ${\rm dim}{\mathbf{\Omega}}_{\{h_i\}}={\rm dim}{\mathbf \Theta}_{\{h_i\}}$, {\emph{a temporal network is structurally reachable, if and only if it is structurally controllable}}.
		
		We remark that the structural reachability notion in Definition \ref{def-structural} is consistent with that in \cite{li2017fundamental,hou2016structural}. They all defined the structural controllability on a specified temporal sequence but did not justify the genericity w.r.t. the duration $\{h_i\}$. As mentioned earlier, since the notion reachability of a switched system based on a single switching signal and that based on the union of all possible switching signals are equivalent, Definition \ref{def-structural} is also consistent with the structural controllability notion of switched systems in \cite{LiuStructural}.
		Similar to Definition \ref{def-structural}, we define the structural overall controllability as follows. 
		\begin{definition}[structural overall reachability] \label{def-structural-2}
			Given the sparsity patterns of $(A_i,B_i)|_{i=1}^N$, we say system (\ref{temporal-plant}) is structurally overall reachable (resp. structurally overall controllable), if there is a realization of $(A_i,B_i)|_{i=1}^N$ that is overall reachable (resp. overall controllable).
		\end{definition}
		
		Notably,  the overall reachability is not necessarily generic w.r.t. $\phi$. To see this, consider a temporal network with $N=3$ and $n=2$. The subsystem parameters are
		{\small $$A_1=\left[\begin{array}{cc}0 & 0 \\
			0 & 0 
			\end{array}\right],B_1=\left[\begin{array}{c}1\\
			0
			\end{array}\right],A_2=\left[\begin{array}{cc}0 & 0 \\
			a_{21} & 0 
			\end{array}\right]$$
			$$B_2=\left[\begin{array}{c}0 \\
			0 
			\end{array}\right],A_3=\left[\begin{array}{cc}0 & a_{12}\\
			0 & 0 
			\end{array}\right],B_3=\left[\begin{array}{c}0\\
			0
			\end{array}\right],$$ with $a_{21}$ and $a_{12}$ indeterminate. 
		}Simple calculations show that the overall reachable set
		$$\begin{array}{c} {\mathbf{\Omega}}=\bigcup_{h_2,h_3>0}e^{A_3h_3}e^{A_2h_2}{\rm Im} B_1\\=\left\{\left[\begin{array}{c}x_1(1+h_2h_3a_{12}a_{21})\\
		x_1h_2a_{21}
		\end{array}\right]:x_1\in {\mathbb{R}}, h_2,h_3>0\right\}.\end{array}$$
		It can be validated that if $a_{12}a_{21}<0$, then ${\mathbf{\Omega}}={\mathbb R}^2$; otherwise, ${\mathbf{\Omega}}\ne {\mathbb R}^2$. Note neither $\{(a_{12},a_{21})\in {\mathbb R}^2: a_{12}a_{21}<0\}$ nor its complement in ${\mathbb R}^2$ is a set of zero measure. Hence, the overall reachability is not generic w.r.t. $(a_{12},a_{21})$. 
		Moreover, the associated reachable set on $\{h_i\}$ is ${\bf \Omega}_{\{h_i\}}={\rm span}[1+h_2h_3a_{12}a_{21},h_2a_{21}]^{\intercal}$, which is a subspace of dimension $1$ for all $a_{12},a_{21}$, and positive $h_2,h_3$. This indicates that for a temporal network, in contrast to the switched systems, {\emph{the overall reachability does not necessarily implies the existence of a temporal sequence on which the system is reachable, and the structural reachability is not necessarily equivalent to the structural overall reachability}}. 
		\subsection{Characterizing the minimal subspace containing ${\mathbf{\Omega}}$} \label{sec-minimal-subspace}
		\begin{proposition} \label{temporal-full-rank}
			Define the subspace ${ \bf \bar \Omega}$ as
			\begin{equation} \label{temporal-form-3}
			{ \bf \bar \Omega}= \sum \limits_{{\tiny k=1,...,N}}^{\tiny j_N,...,j_k=0,...,n-1} A_{N}^{j_N}A_{N-1}^{j_{N-1}}\cdots A_{k+1}^{j_{k+1}}A_{k}^{j_k}{\rm Im} B_{k}.
			\end{equation}
			Then, ${\bf \bar \Omega}$ is the minimal subspace containing the overall reachable set ${\bf \Omega}$. 	
		\end{proposition}
		To prove Proposition \ref{temporal-full-rank}, we need the following lemma.
		\begin{lemma} \citep[Lemma 1]{Z.S2002Controllability} \label{foundamental-set-lemma}
			For any matrix $A\in {\mathbb R}^{n\times n}$ and subspace ${\bf V}\subseteq {\mathbb R}^n$, the following relation holds for almost all $h_1,h_2,...,h_N\in {\mathbb R}$
			$$e^{Ah_1}{\bf V}+e^{Ah_2}{\bf V}+\cdots+e^{Ah_N}{\bf V}= \Gamma_A {\bf V}, $$
			where $\Gamma_A {\bf V}=\sum _{i=0}^{n-1} A^i {\bf V}$.
		\end{lemma}
		
		{\bf Proof of Proposition \ref{temporal-full-rank}:} By the Cayley-Hamilton theorem and the Taylor expansion of $e^{A_ih_i}$ \cite{R.A.1991Topics}, given any $A_i\in {\mathbb R}^{n\times n}$ and $h_i>0$, there
		exist $c_0,c_1,...,c_{n-1}\in {\mathbb R}$ ($c_0,c_1,...,c_{n-1}$ are functions of $h_i$), such that
		\begin{equation} \label{matrix-expansion}
		e^{A_ih_i}=\sum \nolimits_{j=0}^{n-1}c_jA^j.
		\end{equation}
		By substituting (\ref{matrix-expansion}) into (\ref{reachable-set}) for $i=2,...,N$, it follows that ${\bf \Omega}\subseteq \langle A_N|B_N \rangle+\sum_{j=2}^{N}\Gamma_{A_N}\cdots (\Gamma_{A_{j}}\langle A_{j-1}|B_{j-1} \rangle)= {\bf \bar \Omega}$.
		
		
		On the other hand, Let $h_{21},...,h_{2n}$,...,
		$h_{N1},...,h_{Nn}$ be $n(N-1)$ {\emph{positive}} values. Let ${\bf \Omega}_1\doteq \left \langle A_1|B_1 \right \rangle$.
		By Lemma \ref{foundamental-set-lemma}, for almost all $h_{21},...,h_{2n}\in {\mathbb{R}}^+$, we have
		$$\left \langle A_2|B_2\right \rangle +e^{A_2h_{21}}{\bf \Omega}_1+\cdots+e^{A_2h_{2n}}{\bf \Omega}_1\!=\!\left \langle A_2|B_2\right \rangle+ \Gamma_{A_2} {\bf \Omega}_1\!\doteq\! {\bf \Omega}_2.$$
		Similarly, for almost all $h_{31},...,h_{3n}\in {\mathbb{R}}^+$, we have
		$$\left \langle A_3|B_3\right \rangle +e^{A_3h_{31}}{\bf \Omega}_2+\cdots+e^{A_3h_{3n}}{\bf \Omega}_2\!=\!\left \langle A_3|B_3\right \rangle+ \Gamma_{A_3} {\bf \Omega}_2\!\doteq \!{\bf \Omega}_3.$$
		By repeating the above procedure, we have for almost all $h_{N1},...,h_{Nn}\in {\mathbb{R}}^+$,
		\begin{equation} \label{V_N}
		\begin{array}{c}
		\left \langle A_N|B_N\right \rangle +e^{A_Nh_{N1}}{\bf \Omega}_{N-1}+\cdots+e^{A_Nh_{Nn}}{\bf \Omega}_{N-1} \\ = \left \langle A_N|B_N\right \rangle+ \Gamma_{A_N} {\bf \Omega}_{N-1}\doteq {\bf \Omega}_N.
		\end{array}
		\end{equation}
		It is obvious that ${\bf \Omega}_N={\bf \bar \Omega}$. Let us substitute the expressions of ${\bf \Omega}_{N-1},...,{\bf \Omega}_{1}$ into (\ref{V_N}). We get, for almost all $h_{21},...,h_{2n}$,...,$h_{N1},...,h_{Nn}>0$, the following holds
		\begin{equation} \label{equality-set} \begin{aligned}
		& \langle A_N|B_N\rangle + \sum_{i_N=1,...,n}e^{A_Nh_{Ni_N}}\langle A_{N-1}|B_{N-1}\rangle+\\
		& \sum_{i_{N},i_{N-1}=1,...,n}e^{A_Nh_{Ni_N}}e^{A_{N-1}h_{N-1,i_{N-1}}}\langle A_{N-2}|B_{N-2} \rangle+  \\
		& \cdots + \sum_{i_{N},...,i_{2}=1,...,n}e^{A_Nh_{Ni_N}}\cdots e^{A_{2}h_{2,i_{2}}}\langle A_{1}|B_{1} \rangle\\
		& ={\bf \Omega}_N.
		\end{aligned}
		\end{equation} Note each adding item in the left-hand side of (\ref{equality-set}) is a subset of ${\bf \Omega}$.
		This indicates, we can at least find $n$ points $x_1,...,x_n$ from ${\bf \Omega}$, such that ${\rm span}(x_1,...,x_n)={\bf \Omega}_N$ (note that ${\rm dim}\ {\bf \Omega}_N\le n$). This indicates there are $n$ points $x'_1,...,x'_n$ in ${\bf \Omega}$ such that ${\bf \Omega}_N\subseteq {\rm span}(x'_1,...,x'_n)$.
		Combining the fact that ${\bf \Omega}\subseteq {\bf \bar \Omega}$ and $ {\bf \Omega}_N = {\bf \bar \Omega}$, we get that  ${\bf \bar \Omega}$ is the minimal subspace containing ${\bf \Omega}$.  \hfill $\square$
		
		Note ${\mathbf{\Omega}}_{\{h_i\}}\subseteq \bar {\mathbf{\Omega}}$ for any $\{h_i\}_{i=1}^N$. Therefore, a necessary condition for the reachability of (\ref{temporal-plant}) on a single sequence is $\bar {\mathbf{\Omega}}={\mathbb{R}}^n$.  An immediate corollary of Proposition \ref{temporal-full-rank} is as follows, which indicates
		the subspace $\bar {\mathbf{\Omega}}$ is helpful to characterize the vector directions in the set ${\mathbf{\Omega}}$. 
		\begin{corollary} For system (\ref{temporal-plant}), it holds
			$$\max_{x_1,...,x_n\in {\mathbf{\Omega}}} {\rm rank}[x_1,...,x_n]={\rm dim}\bar {\mathbf{\Omega}}.$$
		\end{corollary}
		
		The proof of Proposition \ref{temporal-full-rank} indicates an iterative geometric expression of $\bar {\mathbf{\Omega}}$:
		$$\begin{aligned}
		{\mathbf{\Omega}}_1&=\langle A_1|B_1 \rangle, {\mathbf{\Omega}}_2=\langle A_2|B_2\rangle+\Gamma_{A_2} {\bf \Omega}_1,\cdots\\
		{\mathbf{\Omega}}_i&=\langle A_i|B_i\rangle+\Gamma_{A_{i}} {\bf \Omega}_{i-1}, i=2,...,N.
		\end{aligned}$$ Then, $\bar {\mathbf{\Omega}}={\mathbf{\Omega}}_N$. Define a matrix ${\cal R}$ as
		\begin{equation}\label{R_def}
		{\cal R}\doteq {\bf row}\left\{ A_{N}^{j_N}A_{N-1}^{j_{N-1}}\cdots A_{k}^{j_k} B_{k}|_{k=1,...,N}^{j_N,...,j_k=0,...,n-1}\right\}.
		\end{equation} It follows that $\bar {\mathbf{\Omega}}$ can be algebraically expressed as $\bar {\mathbf{\Omega}}= {\rm span}{\cal R}$. Given the sparsity patterns of $(A_i,B_i)|_{i=1}^N$, since each entry of ${\cal R}$ is a polynomial of $\phi$, it is easy to see that the rank of ${\cal R}$ is generic w.r.t. the values of $\phi$, so is the dimension of $\bar {\mathbf \Omega}$. As above, we denote those generic numbers by ${\rm grank}{\cal R}$ and ${\rm gdim} {\bar {\mathbf{\Omega}}}$, respectively. We remark that \cite{conner1987structure} has decomposed the reachable set of a discrete-time switched system into the union of some `maximal' subspaces. However, such a decomposition is not always possible and is not suitable for the generic analysis.  


		

		Parallel to Proposition \ref{temporal-full-rank}, for the overall controllable set ${\mathbf{\Theta}}$, we have the following result.
		\begin{corollary}
			Define the subspace ${ \bf \bar \Theta}$ as
			\begin{equation} \label{temporal-form-4}
			{ \bf \bar \Theta}= \sum \limits_{{\tiny k=1,...,N}}^{\tiny j_1,...,j_k=0,...,n-1} A_{1}^{j_1}A_{2}^{j_{2}}\cdots A_{k}^{j_k}{\rm Im} B_{k}.
			\end{equation} Then, ${ \bf \bar \Theta}$ is the minimal subspace that contains ${\mathbf \Theta}$.
		\end{corollary}

		%
		
		
		From now on, we focus on the reachability and reachable set, and as discussed above, all these results can be either directly or indirectly extended to their controllability counterparts (thus omitted). Particularly, results for controllability can be obtained by just revising the temporal order of subsystems (i.e., changing the temporal order $(1,2,...,N)$ to $(N,N-1,...,1)$) from the reachability counterparts. Results on the dimension of the reachable subspace are directly valid for that of the controllable subspace.
		
		\section{Verifying structural (overall)  reachability  is `hard'} \label{sec-4}
		In this section, we demonstrate that the following problems associated with the temporal network (\ref{temporal-plant}): \begin{itemize}
			\item[(1)] verifying structural reachability ;
			\item[(2)] verifying structural overall reachability;
			\item[(3)] computing ${\rm gdim}{\mathbf \Omega}_{\{h_i\}}$;
			\item[(4)] computing ${\rm gdim}\bar {\mathbf \Omega}$, \end{itemize} are all at least as hard as the STCP, for any $N\ge 2$. This implies that finding verifiable criteria for structural (overall)  reachability of temporal networks may be unlikely, given that the latter problem is believed to be so. To this end, the definition of structural target controllability is introduced first. Note for an $n\times m$ matrix $M$, $S\subseteq \{1,...,n\}$, we use $M(S,:)$ to denote the matrix consisting of rows of $M$ indexed by $S$.
		
		\begin{definition} [Structural target controllability]\cite{gao2014target,czeizler2018structural}\label{target-control-def}
			Given structured matrices $A$ of size $n\times n$, $B$ of size $n\times m$, and a target set $T\subseteq \{1,...,n\}$, $(A,B,T)$ is said to be structurally target controllable, if there exists a realization $(\tilde A, \tilde B)$ of $(A,B)$, such that the rows of $C_{\tilde A,\tilde B}\doteq [\tilde B,\tilde A\tilde B,...,\tilde A^{n-1}\tilde B]$ indexed by $T$
			have full row rank, i.e., ${\rm rank}C_{\tilde A,\tilde B}(T,:)=|T|$. The STCP is the problem of verifying whether a given $(A,B,T)$ is structurally target controllable.
		\end{definition}
		
		
		Note structural target controllability of $(A,B,T)$ is equivalent to that ${\rm grank}C_{A,B}(T,:)=|T|$ \cite{murota1990note}. Finding complete characterizations for the structural target controllability is a long-standing open issue \cite{murota1990note}. It has been believed that there are no {\emph{deterministic}} necessary and sufficient conditions that can be verified in polynomial time (called verifiable conditions) for the structural target controllability; see \cite{czeizler2018structural,li2020structural} for details. 
		
		\begin{theorem}\label{hardness-temporal} 
			Verifying the structural reachability and the structural overall reachability of (\ref{temporal-plant}), and determining the generic dimension of ${\mathbf{\Omega}}_{\{h_i\}}$, are all at least as hard as STCP for any $N\ge 2$.
		\end{theorem}
		
		
		
		\begin{proof}
			Consider a structured matrix pair, say $(A_1,B_1)$, where $A_1$ and $B_1$ are of size $n\times n$ and $n\times m$ respectively. Consider a target $T\subseteq \{1,...,n\}$, and suppose $|T|<m$. For a given constant integer $N$, where $2\le N \le n-|T|$,  partition the set $\bar T\doteq \{1,...,n\}\backslash T$ into $N-1$ disjoint subsets $S_1,...,S_{N-1}$, $S_j\subseteq \bar T$ for $j=1,...,N-1$.  Construct a temporal network with $N$ subsystems $(A_i, B_i)$, $i=1,...,N$. For $i=2,...,N$, let $A_i=0_{n\times n}$, $B_i=I_{S_{i+1}}$, where $I_{S_{i}}$ denotes the sub-columns of the identity matrix $I$ indexed by $S_{i}$ (the identity matrix is considered as a structured diagonal matrix with nonzero diagonal entries).
			
			Based on the above constructions, it can be seen that the associated reachability matrix
			\begin{equation}\label{hard-reachability}
			{\cal C}=[I_{S_{N-1}}, I_{S_{N-2}}, \cdots, I_{S_1}, C_1],
			\end{equation}
			where $C_1$ is the controllability matrix of $(A_1,B_1)$. And,
			$${\bf \Omega}={\bf \Omega}_{\{h_i\}} = {\rm span}{\cal C}.$$
			By the structure of $I_{S_i}$, it has ${\rm grank} {\cal C}= |\bar T|+ {\rm grank} {C_1}(T,:)$. Therefore, ${\rm grank} {\cal C}=n$, if and only if ${\rm grank} {C_1}(T,:)=|T|$, i.e., $(A_1,B_1,T)$ is structurally target controllable. From the fact that the structural reachability is equivalent that ${\rm gdim}{{\mathbf{\Omega}}_{\{h_i\}}}={\rm grank} {\cal C}=n$, the proposed statements follow immediately.
		\end{proof}
		
		Theorem \ref{hardness-temporal} demonstrates that finding complete characterizations for the structural (overall)  reachability of temporal networks, even with $N=2$ subsystems, is at least as difficult as STCP, which is still unsolved even without the polynomial-time complexity requirement \cite{czeizler2018structural,murota1990note}. Given this, it appears unlikely to obtain verifiable conditions for structural (overall)  reachability of temporal networks. Hence, claims in some literature, such as \cite{hou2016structural}, that such conditions have been achieved may not hold in general. 
		
		\begin{corollary}\label{hard-minimal-space}
			Determining the generic dimension of $\bar {\bf \Omega}$ for any $N\ge 2$ is at least as hard as STCP.
		\end{corollary}
		
		\begin{proof}
			Consider the temporal network constructed in the proof of Theorem \ref{hardness-temporal}. The associated linear space
			$$\bar {\bf \Omega}={\rm Im} I_{S_{N-1}}+{\rm Im} I_{S_{N-2}}+\cdots+ {\rm Im} I_{S_{1}}+\langle A_1|B_1\rangle,$$
			where all the parameters are defined in the same way as therein.
			Thus, ${\rm gdim}\bar {\bf \Omega}=|\bar T|+{\rm grank}C_1(T,:)$. As a result, determining whether ${\rm gdim}\bar {\bf \Omega}=n$ is equivalent to verifying whether ${\rm grank}C_1(T,:)=|T|$, i.e.,
			verifying the structural target controllability of $(A_1,B_1,T)$. This leads to the required statement.
		\end{proof}
		
		Due to the `hardness' of determining the generic dimensions of ${\mathbf{\Omega}}_{\{h_i\}}$ and $\bar {\mathbf{\Omega}}$, in the sequel, we provide some efficiently computable upper and lower bounds for them.
		
		\section{Graph-theoretic upper/lower bounds  of {\small {${\rm gdim}{\mathbf{\Omega}}_{\{h_i\}}$}}} \label{sec-5} 
		In this section, we give a graph-theoretic upper bound and a lower one for the generic dimension of ${\mathbf{\Omega}}_{\{h_i\}}$. As a by-product, we prove Ezzine \& Haddad's conjecture \cite{ezzine1989controllability} for $N=2$ and disprove it for $N\ge 3$. 
		
		Some basic notions of graph theory are introduced, which are needed for the rest of this paper. Let ${\cal G}=(V,E)$ be a directed graph (digraph) with vertex set $V$ and edge set $E\subseteq V\times V$. A subgraph ${\cal G}_s=(V_s,E_s)$ of ${\cal G}$ is a graph such that $V_s\subseteq V$ and $E_s\subseteq E$, and is called a subgraph induced by $V_s$ if $E_s=V_s\times V_s \cap E$. We say ${\cal G}_s$ covers $V_s'\subseteq V$, if $V_s'\subseteq V_s$.  A path $P$ from $i_1$ to $i_k$ in ${\cal G}$ is a sequence of edges $(i_1,i_2)$, $(i_2,i_3)$,...,$(i_{k-1},i_k)$ with $(i_j,i_{j+1})\in E$, $j=1,...,k-1$, which is also denoted by $P=(i_1,i_2,...,i_k)$. Note we do {\emph{not}} require that vertices $i_1,...,i_k$ are distinct. We call vertex $i_1$ the tail (i.e., initial vertex) of path $P$, denoted by ${\rm tail}(P)$, and vertex $i_k$ the head (i.e., terminal vertex) of path $P$, denoted by ${\rm head}(P)$. If $i_1$ and $i_k$ are the only repeated vertices in $P$, then $P$ is a {\emph{cycle}}. The length of a path is the edges it contains. A multigraph is a graph that allows multiple edges (also called parallel edges) between the same pair of vertices. 
		
		If vertices of a graph ${\cal G}$ can be partitioned into two parts (called bipartitions) such that no edge has its two end vertices within the same part, ${\cal G}$ is called a
		bipartite graph, and denoted by ${\cal G}=(V_1,V_2,E)$, with $V_1,V_2$ its bipartitions and $E$ the edges. A matching in ${\cal G}$ is a set of edges so that no two edges share the same end vertices. A maximum matching of ${\cal G}$ is a matching with the largest number of edges. Assign a non-negative weight to each edge of ${\cal G}$. Then, a maximum weighted matching of ${\cal G}$ is a matching with the largest sum of weights of edges it contains.  For a structured matrix $M$, its associated bipartite graph is defined as ${\cal B}(M)=(R,C,E_M)$, with $R$ and $C$ the row and column indices of $M$, and $E_M=\{(i,j): i\in R, j\in C, M_{ij}\ne 0\}$. It is known that the maximum size of a matching in ${\cal B}(M)$ equals ${\rm grank} M$ \citep[Proposition 2.1.12]{Murota_Book}.
		
		For the $i$th subsystem of system (\ref{temporal-plant}), $i=1,...,N$, let ${\cal G}_i=(V_{A_i}\cup V_{B_i}, E_{A_i}\cup E_{B_i})$ be its {\emph{system digraph}}, with state vertices $V_{A_i}=\{v^i_1,...,v^i_n\}$, input vertices $V_{B_i}=\{v^i_{n+1},...,v^i_{n+m_i}\}$, and edges $E_{A_i}=\{(v^i_k,v^i_j): A_{i,jk}\ne 0\}$, $E_{B_i}=\{(v^i_{n+k},v^i_{j}): B_{i,jk}\ne 0\}$.  Let $X=\{v_1,...,v_n\}$ be the vertex set representing the states of the whole temporal network (\ref{temporal-plant}). In the sequel, $V_{A_i}|_{i=1}^N$ can be seen as $N$ copies of $X$. Let the digraph ${\cal G}(A_i)\doteq (V_{A_i},E_{A_i})$. A state vertex $v\in V_{A_i}$ is said to be {\emph{input-reachable}} in ${\cal G}_i$, if there is a path from some $u\in V_{B_i}$ to $v$ in ${\cal G}_i$. A {\emph{stem}} is a path from an input vertex with no repeated vertices in~${\cal G}_i$.

		\subsection{Upper bound of ${\rm gdim}{\mathbf{\Omega}}_{\{h_i\}}$ based on the CDG} \label{upper-bound-sec}

		For the $i$th subsystem $(A_i,B_i)$, construct a digraph $\bar {\cal G}_i=(\bar V_{A_i}\cup \bar V_{B_i},\bar E_i)$, $i=1,...,N$, where the vertex set $\bar V_{A_i}=\{v^i_{jt}:j=1,...,n,t=1,...,n\}$, $\bar V_{B_i}=\{v^i_{n+j,t}: j=1,...,m_i, t=0,...,n-1\}$, and the edge set $\bar E_i=\{(v^i_{jt},v^i_{k,t+1}): A_{i,kj}\ne 0, t=1,...,n-1\}\cup \{(v^i_{n+j,t},v^i_{k,t+1}): B_{i,kj}\ne 0, t=0,...,n-1\}$. Such an acyclic digraph is called the {\emph{dynamic graph}} of system $(A_i,B_i)$ in \cite{poljak1990generic,murota1990note}.
		Define the edge set \begin{equation} \label{connect-E}
		\bar E_{i-1,i}=\left\{(v^{i-1}_{kn},v^i_{jn}): \exists \ {\rm a \ path \ from}\ v^i_k \ {\rm to } \ v^i_j \ {\rm in} \ {\cal G}(A_i)\right\},
		\end{equation}for $i=2,...,N$. Note it always has $(v^{i-1}_{kn},v^i_{kn})\in \bar E_{i-1,i}$, $\forall k=1,...,n$.
		Connect those ${\bar {\cal G}}_i|_{i=1}^N$ by adding edges $\bar E_{i-1,i}|_{i=2}^N$, and we get the digraph $\bar {\cal G}$, which is called the {\emph{cascaded dynamic graph}} (CDG) of the temporal network (\ref{temporal-plant}). Let $\bar V_B=\bigcup_{i=1}^N \bar V_{B_i}$, and $\bar V_{A_N,n}=\{v^N_{1n},...,v^N_{nn}\}$.  Assign weights to the edges of $\bar {\cal G}$ as
		$w(e)=A_{i,kj}$ if $e=(v^i_{jt},v^i_{k,t+1})$, and $w(e)=B_{i,kj}$ if $e=(v^i_{n+j,t},v^i_{k,t+1})$.  The weights of edges in $\bar E_{i-1,i}$ will be defined in the sequel.
		
		We adopt some terminologies from \cite{murota1990note}.
		A collection $L=(p_1,...,p_k)$ of vertex-disjoint paths in $\bar {\cal G}$ is called a linking. The size of a linking $L$ is the number of paths it contains. We call $L$ a $S-T$ linking, if $\{{\rm tail}(p_i):i=1,...,k\}\subseteq S$, and $\{{\rm head}(p_i): i=1,...,k\}\subseteq T$. When $L$ contains only one path, we may call it a $S-T$ path.  We are interested in the linkings from $\bar V_B$ to $\bar V_{A_N,n}$, i.e., $S-T$ linkings when $S\subseteq \bar V_B$ and $T\subseteq \bar V_{A_N,n}$. For such a linking $L$, define $w(p_i)$ as the product of weights of all individual edges in each path $p_i$ of $L$, $i=1,...,k$, and $w(L)=w(p_1)w(p_2)\cdots w(p_k)$.
		Let $\prec$ be a fixed order of vertices in $\bar V_B$ and $\bar V_{A_N,n}$. For a linking $L=(p_1,...,p_k)$, let $s_1\prec s_2\prec\cdots \prec s_k$ and $t_1\prec t_2\prec\cdots \prec t_k$ be respectively
		the tails and heads of $L$. Moreover, suppose $s_{\pi{(i)}}$ and $t_{i}$ are the tail and head of the path $p_i$, $i=1,...,k$. Then, we obtain a permutation $\pi\doteq (\pi(1),...,\pi(k))$ of $1,2,...,k$, and denote its sign by ${\rm sign}(\pi)\in \{1,-1\}$. The sign of the linking $L$ is defined as ${\rm sign}(L)={\rm sign}(\pi)$.
		To illustrate the above construction, we provide next an example.
		
		\begin{example}Consider the temporal network with $n=3$, $N=2$ shown in Fig. \ref{first-example}. The subsystem digraphs ${\cal G}_1$ and ${\cal G}_2$, and the associated dynamic graphs $\bar{\cal G}_1$, $\bar{\cal G}_2$, and $\bar{\cal G}$ are given thereof.  In the CDG $\bar{\cal G}$, upon letting $p_1=(v_{42}^1,v_{13}^1,v_{13}^2)$, $p_2=(v^1_{41},v^1_{12},v^1_{23},v^2_{33})$, $L=(p_1,p_2)$ is a $\{v^1_{42},v^1_{41}\}-\{v^2_{13},v^2_{33}\}$ linking (bold lines in Fig. \ref{first-example}), which is of the maximum size in all linkings from $\bar V_B$ to $\bar V_{A_{22}}$. If we presume a fixed order $v^1_{40}\prec v^1_{41}\prec v^1_{42}$, $v^2_{13}\prec v^2_{23}\prec v^2_{33}$, then ${\rm sign}(L)=-1$.
		\end{example}

		Our idea to bound ${\rm gdim}{\bf \Omega}_{\{h_i\}}$ is introducing the structured matrix $M_i$ that has the same sparsity pattern as $e^{A_ih_i}$ for almost all $h_i\in {\mathbb R}$ and $\phi\in {\mathbb R}^d$, $i=1,...,N$. ~Let \begin{equation} \label{key-surrogate}
		\bar {\cal C}\doteq [C_N, M_NC_{N-1},\cdots, M_NM_{N-1}\cdots M_2C_1].
		\end{equation}
		Since $M_i$ ignores the possible parameter dependence among the nonzero entries of $e^{A_ih_i}$, it is obvious ${\rm grank}\bar {\cal C} \ge {\rm grank} {\cal C}$.  The next result gives an upper bound of ${\rm grank} \bar {\cal C}$, thus upper bounding ${\rm gdim}{\bf \Omega}_{\{h_i\}}$.
		

		
		\begin{theorem} \label{upper-bound}
			The generic dimension of the reachable subspace ${\bf \Omega}_{\{h_i\}}$ is no more than the maximum size of a linking from $\bar V_B$ to  $\bar V_{A_N,n}$ in $\bar {\cal G}$.
		\end{theorem} 
		
		\begin{proof} We begin by noting that $e^{A_it}={\cal L}^{-1}((sI-A_i)^{-1})$, where $t$ represents the time variable, $s$ the frequency variable, and ${\cal L}^{-1}(\cdot)$ the point-wise inverse Laplace transform \cite{antsaklis1997linear}. This implies that $e^{A_ih_i}$ exhibits the same sparsity pattern as $(sI-A_i)^{-1}$ for almost all $h_i\in {\mathbb R}$ and $\phi \in {\mathbb R}^d$. Note, $(sI-A_i)^{-1}$ serves as the transfer function matrix from all state variables to themselves. Therefore, $[(sI-A_i)^{-1}]_{kj}\ne 0$ if and only if there exists a path from $v^i_j$ to $v^i_k$ in the digraph ${\cal G}(A_i)$.

			By the definitions of $\bar E_{i-1,i}$ and $M_i$, it follows that $[M_i]_{kj}\ne 0$ implies that $(v^{i-1}_{jn},v^i_{kn})\in \bar E_{i-1,i}$. Let the weight of edge $e=(v^{i-1}_{jn},v^i_{kn})$ in $\bar E_{i-1,i}$ be $w(e)=[M_i]_{kj}$.
			For $I\subseteq \bar V_{A_N,n}$ and $J\subseteq \bar V_B$ with $|I|=|J|=l$, let $\bar  {\cal C}(I,J)$ be the square submatrix of $\bar  {\cal C}$ with rows indexed by $I$ and columns by $J$. The entry at $(v_{in}^N,v^k_{n+j,t})$ of $\bar  {\cal C}$, $v_{in}^N\in I$, $v^k_{n+j,t}\in J$, given by $\bar  {\cal C}(v_{in}^N,v^k_{n+j,t})$, is
			\begin{equation}\label{entry}
			\bar  {\cal C}(v_{in}^N,v^k_{n+j,t})=\sum \limits_{p: \ all \ (v^k_{n+j,t}, v_{in}^N) \ paths \ in \ {\bar {\cal G}}} w(p).
			\end{equation}
			Suppose vertices of $I$ and $J$ have a fixed ordering, such that for every permutation $\pi$ of $1,...,l$, the $i$th element of $I$ is mapped to the $\pi(i)$th element of $J$, $i=1,...,l$.
			By the definition of determinant, we have
			\begin{equation}\label{determinant} \det \bar {\cal C}(I,J)= \sum \limits_{\pi} {\rm sign(\pi)} \prod_{v_{in}^N\in I} \bar  {\cal C}(v_{in}^N,\pi(v_{in}^N)), \end{equation}
			where the summation is taken over all permutations $\pi: I\to J$. 
			
			Next, we invoke a proof technique similar to \cite{poljak1989maximum}. For a given $\pi: I\to J$, let $L'(\pi)=(p_1,...,p_l)$ be a collection of $l$ paths ($l=|J|$), satisfying $\pi({\rm head}(p_i))={\rm tail}(p_i)$, $i=1,...,l$.
			Substituting (\ref{entry}) into (\ref{determinant}) yields{\small
				\begin{equation}\label{determinant-2} \begin{aligned}
				\det\bar {\cal C}(I,J)\!\!&=\!\! \sum \limits_{\pi} {\rm sign}(\pi) \prod_{v_{in}^N\in I}\ \sum \limits_{p: \ (\pi(v_{in}^N), v_{in}^N) \ paths \ in \ {\bar {\cal G}}} w(p) \\
				\!\!&=\!\!\sum  \limits_{\pi} {\rm sign}(\pi) \sum \limits_{L'(\pi)} \ \prod \limits_{p_{\pi(v_{in}^N),v_{in}^N}\in L'(\pi)} w(p_{\pi(v_{in}^N),v_{in}^N}),
				\end{aligned}
				\end{equation}}where $p_{\pi(v_{in}^N),v_{in}^N}$ denotes the weight of a path from $\pi(v_{in}^N)$ to $v_{in}^N$ in $L'(\pi)$, and the sandwiched summation is taken over all possible $L'(\pi)$.
			We observe that if the $l$ paths in $L'(\pi)$ are not vertex-disjoint, then for each term $\prod \nolimits_{p_{i}\in L'(\pi)} w(p_{i})$, we can always find another permutation $\pi'$ and the associated $l$ paths $L'(\pi')$ satisfying  $$\prod \nolimits_{p_{i}\in L'(\pi')} w(p_{i})=\prod \nolimits_{p_{i}\in L'(\pi)} w(p_{i})$$ but ${\rm sign}(\pi')=-{\rm sign}(\pi)$. Indeed, if two paths $p_{\pi(i),i}$ and $p_{\pi(i'),i'}$ intersects at a vertex $v$, let $w$ and $\sigma$ be respectively the $(\pi(i),v)$ path and $(v,i)$ path in $p_{\pi(i),i}$, and $w'$ and $\sigma'$ be the $(\pi(i'),v)$ path and $(v,i')$ path in $p_{\pi(i'),i'}$. Then, we obtain two new paths by connecting $w$ with $\sigma'$ and $w'$ with $\sigma$, and the remaining paths remain unchanged. This generates a new collection of $l$ paths $L'(\pi')$, with $\pi'(i)=\pi(i')$ and $\pi'(i')=\pi(i)$, leading to ${\rm sign}(\pi')=-{\rm sign}(\pi)$, while $\prod \nolimits_{p_{i}\in L'(\pi')} w(p_{i})=\prod \nolimits_{p_{i}\in L'(\pi)} w(p_{i})$. Hence, all collections of $l$ paths $L'(\pi)$ that are not vertex-disjoint will cancel out in (\ref{determinant-2}). Consequently, we have
			\begin{equation} \label{det-important}
			\det \bar {\cal C}(I,J)= \sum \limits_{L: {\rm all \ } (J,I) \ {\rm linkings \ in }\ {\bar {\cal G}}} {\rm sign}(L)w(L).
			\end{equation} To make $\det \bar {\cal C}(I,J)\ne 0$, a size $|I|$ linking must exist. Hence, ${\rm grank}\bar {\cal C}$ is no more than the maximum size of a linking from $\bar V_B$ to $\bar V_{A_{N,n}}$ in $\bar {\cal G}$. The required statement then follows from the fact that ${\rm grank}  {\cal C}\le {\rm grank} \bar {\cal C}$.
		\end{proof}
		\begin{figure}
			\centering
			\includegraphics[width=2.4in]{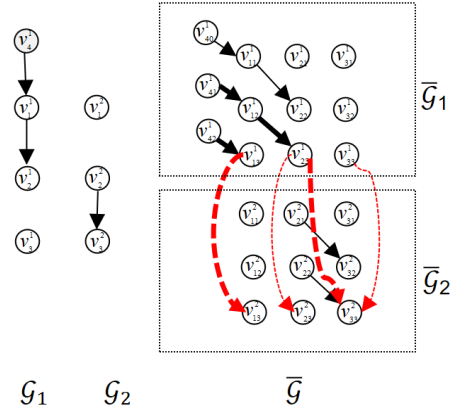}\\
			\caption{A temporal network and the associated graph representations. Dotted red lines represent edges in $\bar E_{1,2}$, while bold ones form a $\bar V_B-\bar V_{A_{22}}$ linking of the maximum size.}\label{first-example}
		\end{figure}
		\begin{remark}
			From \cite{poljak1990generic}, when $N=1$, the value of ${\rm grank}C_1$ is precisely the maximum size of a $\bar V_{B_1}-\bar V_{A_{1,n}}$ link in $\bar {\cal G}$. Therefore, Theorem \ref{upper-bound} extends the dynamic graph presented in \cite{poljak1990generic,murota1990note} from a single system to temporal networks.
		\end{remark} 

		\begin{remark}\label{reason-for-tight}
			Let $L$ be a linking, and let $H_L$ be the multigraph obtained by taking the union of the edges in ${\cal G}_i|_{i=1}^N$ and $\bar E_{i-1,i}|_{i=2}^N$ that correspond to all edges in each individual path of $L$. In particular, an edge $(v^i_{j,t},v^i_{k,t+1})$  in $\bar {\cal G}$ corresponds to $(v^i_{j},v^i_{k})$ in ${\cal G}_i$. The reason why Theorem \ref{upper-bound} may sometimes fail to give the exact value of ${\rm grank}\bar {\cal C}$ is that there may exist two different $(I,J)$ linkings $L$ and $L'$ with the maximum size that produce the same multigraphs $H_L$ and $H_{L'}$ (so that $w(L)=w(L')$) but have opposite signs ${\rm sign}(L)=-{\rm sign}(L')$, making them cancel out in (\ref{det-important}). It becomes clear that only when for every linking $L$ of the maximum size, there is a distinct linking $L'$ satisfying the aforementioned cancellation condition, the maximum size of linkings can mismatch ${\rm grank}\bar {\cal C}$. However, as shown in \cite{murota1990note} for the structural output/target controllability problem, such a situation is rare and requires a carefully designed example. Therefore, for practical systems, Theorem \ref{upper-bound} is highly likely to give an accurate estimate of ${\rm grank}\bar {\cal C}$ \cite{czeizler2018structural,gao2014target}. In fact, the maximum size linking condition has been used as an `almost' necessary and sufficient condition for structural target controllability in \cite{gao2014target}. However, it is important to note that the difference between ${\rm grank}\bar {\cal C}$ and ${\rm grank}{\cal C}$ is still not well understood. Nevertheless, our simulations suggest that for temporal networks with subsystem structures generated at random, it is highly likely that ${\rm grank}\bar {\cal C}$ and ${\rm grank}{\cal C}$ coincide.

		\end{remark}
		
		Computing the maximum size of linkings can resort to the maximum flow algorithm; see \citep[Section 10.2]{generic} for detail. Since $\bar {\cal G}$ has $M\doteq Nn^2+n\sum_{i=1}^Nm_i$ vertices, the maximum size of a linking in $\bar {\cal G}$ could be obtained in time at most $O(M^3)\to O(N^3n^6)$ by \cite{lawler2001combinatorial}.

		\subsection{Lower bound of  the generic dimension of ${\bf \Omega}_{\{h_i\}}$}
		From Proposition \ref{generic-dimension-2}, by taking $h_1=\cdots=h_N=0$, a lower bound of ${\rm gdim} {\bf \Omega}_{\{h_i\}}$ is the generic rank of ${\cal C}_{\rm low}$ with
		$${\cal C}_{\rm low}\doteq [C_N,C_{N-1},\cdots, C_1].$$
		It is worth noting that ${\rm rank}{{\cal C}_{\rm low}}=n$ has been used by many researchers as a sufficient condition for reachability (controllability) of switched/hybrid systems; see \cite{ezzine1989controllability,sun2001reachability}.
		
		Using the similar arguments to the proof of Theorem \ref{hardness-temporal}, it can be seen that {\emph{determining ${\rm grank}{{\cal C}_{\rm low}}$ is again at least as hard as the STCP, for any $N\ge 2$}}. Note in the case specified in the proof of Theorem \ref{hardness-temporal}, ${{\cal C}_{\rm low}}$ is of the same form as (\ref{hard-reachability}).
		
		A natural idea to evaluate ${\rm grank}{\cal C}_{\rm low}$ is to change $\bar E_{i-1,i}$ in (\ref{connect-E}) to
		$$\bar E_{i-1,i}=\{(v^{i-1}_{kn},v^i_{kn}):k=1,...,n\}.$$ Denote the obtained $\bar {\cal G}$ by $\bar {\cal G}'$. Then,
		following Theorem \ref{upper-bound}, we can use the maximum size of a linking in $\bar {\cal G}'$ to estimate ${\rm grank}{\cal C}_{\rm low}$. However, this is only guaranteed to get an upper bound of ${\rm grank}{\cal C}_{\rm low}$, rather than a lower one for ${\rm gdim} {\bf \Omega}_{\{h_i\}}$. To give a lower bound of ${\rm gdim} {\bf \Omega}_{\{h_i\}}$, we resort to the cactus configuration introduced as follows.
		

		\begin{definition}\label{cactus-original} \cite{hosoe1980determination}
			A set $V_i\subseteq X$ is said to be covered by a cactus configuration in ${\cal G}_i$, if 1) every $v\in V'_i \doteq \{v^i_j: v_j\in X\}$ is input-reachable in ${\cal G}_i$, and 2) $V'_i$ can be covered by a collection of disjoint cycles and stems in ${\cal G}_i$. Any subset of a covered set $V_i$ is also covered.
		\end{definition}
		
		\begin{lemma}\cite{hosoe1980determination}\label{subspace-theorem}
			For a single system, say $(A_1,B_1)$, the generic dimension of its reachable subspace equals the maximum size of a set $V_1\subseteq X$ that can be covered by a cactus configuration in ${\cal G}_1$.
		\end{lemma}
		
		\begin{proposition} \label{low-union}
			Let $V_i\subseteq X$ be a set that is covered by a cactus configuration in ${\cal G}_i$, $i=1,...,N$. Then, ${\rm grank} \bar{\cal C}_{\rm low}\ge |\bigcup\nolimits_{i=1}^N V_i|$
		\end{proposition}
		
		\begin{proof}
			We will prove the case $N=2$, and cases $N>2$ can follow a similar manner. Suppose $\bigcup_{i=1}^2V_i=V_1'\cup V_2'$, with $V_1'\subseteq V_1$, $V_2'\subseteq V_2$ and $V_1'\cap V_2'=\emptyset$. Consider the CDG $\bar {\cal G}'$ associated with $[C_2,C_1]$. Let $\bar V^1_{A_2,n}=\{v^2_{jn}: v_j\in V_1'\}$, and $\bar V^2_{A_2,n}=\{v^2_{jn}: v_j\in V_2'\}$. By Lemma \ref{subspace-theorem} and also the proof of \citep[Theorems 1,3]{poljak1989maximum}, there is a linking $L_1$ from some $\bar V'_{B_1}\subseteq \bar V_{B_1}$ to $\bar V^1_{A_2,n}$, $|\bar V'_{B_1}|=|\bar V^1_{A_2,n}|$, which is determined by the cactus configuration covering $V_1'$. By virtue of the vertex-disjointness, upon denoting $H_{L_1}$ as the multigraph obtained from the union of edges in ${\cal G}_1$ that correspond to edges in ${L_1}$, no other $\bar V'_{B_1}-\bar V^1_{A_2,n}$ linking in $\bar {\cal G}'$ can produce the same multigraph $H_{L_1}$ (as well as the same weight $w({L_1})$). Moreover, for any other $\bar V'_{A_2,n}\ne \bar V^1_{A_2,n}$, obviously, there is no other $\bar V'_{B_1}-\bar V'_{A_2,n}$ linking $L'$ with the weight equaling $w(L_1)$. Let $\bar V'_{B_2}$ and $L_2$ be defined similarly. Then, similar properties are true for the $\bar V'_{B_2}-\bar V^2_{A_2,n}$ linking $L_2$. Consequently, $L_1\cup L_2$ is a $\bar V'_{B_1}\cup V'_{B_2}- \bar V^1_{A_2,n}\cup \bar V^2_{A_2,n}$ linking with a unique weight that cannot be canceled out by any other linking from $\bar V'_{B_1}\cup V'_{B_2}$ to $\bar V^1_{A_2,n}\cup \bar V^2_{A_2,n}$ in $\bar {\cal G}'$. By (\ref{det-important}) in the proof of Theorem \ref{upper-bound}, the submatrix consisting of rows of $[C_2,C_1]$ indexed by $V_1'\cup V_2'$ generically has full row rank.
		\end{proof}
		
		Let $U_i$ be the collection of maximal state vertex sets that are covered by a cactus configuration in ${\cal G}_i$.\footnote{A maximal set here means this set cannot be a proper subset of any other set that is covered by a cactus configuration.}  Then, a lower bound of ${\rm grank}{\cal C}_{\rm low}$ is the solution of the following problem:
		$${\rm Problem} \ {\cal P}: \max_{V_i\in U_i} |\bigcup\nolimits_{i=1}^N V_i|.$$ However, since each $U_i$ can contain an exponential number (in $n, m_i$) of elements,
		optimizing problem ${\cal P}$ seems hard due to its combinatorial structure. Even for $N=2$, a brute force may need to check the size of an exponential number of unions ${V_1\cup V_2}$. In the following, we provide a greedy heuristic for problem ${\cal P}$ in polynomial time with provable approximation guarantees.
		
		%

		\begin{algorithm}[H] 
			{{{
						\caption{: A greedy heuristic for problem ${\cal P}$}
						\label{alg1} 
						\begin{algorithmic}[1]
							\STATE {Set $E_0=X$, ${\cal I}=\{1,...,N\}$, $\tilde V_0=\emptyset$, $k=0$;
							}
							\WHILE {${\cal I}\ne \emptyset$ and $\tilde V_{k}\backslash \tilde V_{k-1}\ne \emptyset$ ({\small {suppose the first iteration with $k=0$ is always executed}})} 
							\STATE Determine the set $V_i\subseteq E_k$ with the maximum size that is covered by a cactus configuration in ${\cal G}_i$, for each $i\in {\cal I}$.
							\STATE Find $i^*={\arg \max_{i\in {\cal I}} |V_i|}$.
							\STATE Update $\tilde V_{k+1}= \tilde V_{k}\cup V_{i^*}$, ${\cal I}={\cal I}\backslash \{i^*\}$, $E_{k+1}= X \backslash \tilde V_{k+1}$, and $k=k+1$.
							\ENDWHILE
							\STATE Return $\tilde V_{k}$ with size $|\tilde V_k|$.
				\end{algorithmic}}}
			}
		\end{algorithm}
		
		Step 3 of Algorithm \ref{alg1}, i.e., finding a subset with the maximum size from a given set $E_k$ that is covered by a cactus configuration in ${\cal G}_i$, can be implemented using the maximum weighted matching algorithm; see \citep[Theorem 6]{murota1990note} for details.
		
		To describe the approximation bounds of Algorithm \ref{alg1}, we introduce some notations first. Let $\tilde V_{k}$ be the solution returned by Algorithm \ref{alg1}, and $\tilde V^*$ the optimal solution to problem ${\cal P}$. In addition, let $V_{i_1^*}$ be the set selected in the first iteration of Algorithm \ref{alg1}, whose size equals the maximum generic dimension of reachable subspace over all subsystems $(A_i,B_i)|_{i=1}^N$. For $N\ge 2$ and $t\in \{0,1,...,N-2\}$, define the function $f(t)$ as
		\begin{equation} \label{relation0}
		{\small{f(t)=\frac{t}{N-1}|\tilde V^*|+[\frac{N-t-1}{N-1}+\sum_{j=1}^{t}\frac{N-t-1}{N-j}-t]|V_{i_1^*}|.}}
		\end{equation}
		
		\begin{proposition}\label{boud-alg1} For $N\ge 2$, the solution $\tilde V_{k}$ returned by Algorithm \ref{alg1} satisfies
			\begin{equation} \label{relation1} |\tilde V_k|\ge \frac{2N-3}{N(N-1)}|\tilde V^*|.\end{equation}
			In particular, for $N=2$ and $3$, Algorithm \ref{alg1} achieves a $\frac{1}{2}$-approximation. Moreover, for $N\ge 3$,
			\begin{equation} \label{relation2}
			|\tilde V_k|\ge \max\limits_{0\le t\le N-2}f(t).
			\end{equation}
			
		\end{proposition}
		\begin{proof}
			See the appendix.
		\end{proof}
		
		
		\begin{remark}Compared to the conventional set cover problem (i.e., given a collection $V_1,...,V_N$ of subsets of a set, finding $k$ members from $V_1,...,V_N$ such that their union is maximum; see \cite{nemhauser1978analysis}), the difficulty of $\max_{V_i\in U_i}|\bigcup_{i=1}^N V_i|$ lies in that, each member $V_i$ must come from a different set $U_i$. Therefore, unlike the approximation factor (i.e.,$1-1/e$) obtained in the set cover problem \cite{nemhauser1978analysis}, Algorithm \ref{alg1} cannot achieve a better approximation factor than $1/2$ for any $N\ge 2$. This can be seen from the following example. Consider a temporal network consisting of $N$ subsystems, with $U_1=\{\{v_1,v_2,v_3\},\{v_4,v_5,v_6\}\}$, $U_2=\{\{v_4,v_5,v_6\}\}$, and $U_i=\{\emptyset\}$ for $i=3,...,N$. It is easy to see that such a temporal network can exist.  For this temporal network, Algorithm \ref{alg1} may first choose $\{v_4,v_5,v_6\}$ from $U_1$ and return a solution $\{v_4,v_5,v_6\}$ eventually. However, the optimal solution  is $\{v_4,v_5,v_6,v_1,v_2,v_3\}$. In this case, Algorithm \ref{alg1} achieves a $1/2$ approximation factor.
		\end{remark}
		
		\subsection{Relations between  ${\cal C}$ and ${\cal C}_{\rm low}$: on Ezzine \& Haddad's conjecture}
		
		Ezzine \& Haddad \cite{ezzine1989controllability} conjectured that given $(A_i,B_i)|_{i=1}^N$, it is `almost always' true that ${\rm rank} {\cal C}={\rm rank} {\cal C}_{\rm low}$ for $h_1,...,h_N$ $>0$\footnote{Here, `almost always true' is understood as `true for almost all $h_1,...,h_N>0$'.}, with ${\cal C}$ given by (\ref{reachability-matrix}).
		Using the graph-theoretic upper and lower bounds established in this section, we have the following result regarding this conjecture.
		
		\begin{corollary} \label{conjecture-ans}
			Ezzine \& Haddad's  conjecture \cite{ezzine1989controllability} is true for $N=2$ but not for any $N\ge 3$.
		\end{corollary}
		
		\begin{proof} Consider $N=2$. Given $(A_i,B_i)|_{i=1}^2$, for any $h_2>0$, we have ${\rm rank}[C_2, e^{A_2h_2}C_1]={\rm rank}e^{-A_2h_2}[C_2, e^{A_2h_2}C_1]$\\$={\rm rank}[e^{-A_2h_2}C_2, C_1]$. By the Cayley-Hamilton theorem, there are $c_0,c_1,...,c_{n-1}\in {\mathbb R}$ such that $e^{-A_2h_2}=\sum_{i=0}^{n-1}c_iA_2^i$. By virtue of this, ${\rm span} e^{-A_2h_2}C_2\subseteq {\rm span} C_2$. Therefore, $${\rm rank}[e^{-A_2h_2}C_2, C_1]\le {\rm rank}[C_2,C_1].$$On the other hand, by Proposition \ref{generic-dimension}, for almost all $h_2>0$,
			$${\rm rank}[e^{-A_2h_2}C_2, C_1]\ge {\rm rank}[e^{-A_2h_2}C_2,C_1]_{h_2=0}={\rm rank}[C_2,C_1].$$
			Taken together, we have ${\rm rank}[C_2, e^{A_2h_2}C_1]={\rm rank}[C_2,C_1]$ for almost all $h_2>0$ and any numerical $(A_i,B_i)|_{i=1}^2$, proving Ezzine \& Haddad's  conjecture with $N=2$.
			
			To disprove Ezzine \& Haddad's  conjecture with $N=3$, we just need to construct a counter-example. Consider a temporal network with $N=3$ subsystems, whose subsystem digraphs are respectively ${\cal G}_1$, ${\cal G}_2$, and ${\cal G}_1$ shown in Fig. \ref{first-example}. That is, $(A_3,B_3)$ of this temporal network is structurally equivalent to $(A_1,B_1)$. For this system, since $C_2=0$, it turns out ${\rm grank} {\cal C}_{\rm low}={\rm grank} C_1 =2$. However, by assigning random values to $(A_i,B_i)$, $i=1,...,3$, it can be checked for almost all $h_2,h_3>0$, ${\rm rank} {\cal C}=3$ (in fact, one can check that ${\rm grank}{\bar {\cal C}}=3$ from Fig. \ref{switch-design} of Section \ref{sec-application}; see  Example \ref{switched-example}). It turns out that for almost all realizations of this temporal network, Ezzine \& Haddad's  conjecture is not true. 
			
			For $N\ge 4$, one just needs to construct a temporal network with its $N$ subsystem digraphs being ${\cal G}_1$, ${\cal G}_2$, ${\cal G}_1$, ${\cal G}_2$,..., where the last one is ${\cal G}_1$ if $N$ is odd, and ${\cal G}_2$ otherwise. It is easy to check ${\rm grank} {\cal C}_{\rm low}=2$ while ${\rm grank} {\cal C}=3$, disproving Ezzine \& Haddad's  conjecture with $N\ge 3$.
		\end{proof}
		From Corollary \ref{conjecture-ans}, ${\rm grank} {\cal C}$ and ${{\rm grank}} {\cal C}_{\rm low}$ always coincide for $N=2$, while not for $N\ge 3$. As a result, for $N=2$, one can alternatively use the maximum size of linkings in $\bar {\cal G}'$ (the CDG associated to $[C_2,C_1]$) to obtain an upper bound for ${\rm gdim}{\mathbf{\Omega}}_{\{h_i\}}$.

		
		
		
		\section{Graph-theoretic upper/lower bounds of ${\rm gdim}{\bar {\bf  \Omega}}$} \label{sec-6}
		In this section, we give graph-theoretic upper and lower bounds for ${\rm gdim}{\bar {\bf \Omega}}$. In obtaining these bounds, two new graph-theoretic notions are introduced, namely, the multi-layer dynamic graph and the temporal cactus. 
		

		
		\subsection{Upper bound of the generic dimension of $\bar {\bf \Omega}$}
		Recall $\bar {\bf \Omega} $ in (\ref{temporal-form-3}) is the summation of $\sum\nolimits_{k=1}^N n^{N-k+1}$ $>n^N$ items $A_{N}^{j_N}\cdots A_{k}^{j_k} {\rm Im}B_{k}$, whose number increases exponentially with $N$. This makes the conventional (dynamic) graph representation of $\bar {\bf \Omega}$ (and its associated matrix ${\cal R}$ defined in (\ref{R_def})) has an exponential number of vertices in $n$ and $N$. To handle this situation, inspired by \cite{Z.S2002Controllability}, we first introduce another iterative geometric expression of $\bar {\bf \Omega}$.
		
		
		Consider the nested subspaces as
		\begin{equation} \label{subspaces_series_cons}
		\begin{array}{c}
		{\bf \Phi}_{01}={\rm Im}\ B_1, {\bf \Phi}_{02}={\rm Im} \ B_2,\cdots, {\bf \Phi}_{0N}= {\rm Im} \ B_N,\\
		{\bf \Phi}_{ij}=A_j \sum \nolimits_{k=1}^j {\bf \Phi}_{i-1,k}, j=1,2,\cdots,N, i=1,2,\cdots.
		\end{array}
		\end{equation}
		Let ${\bf W}_i=\sum_{k=0}^{i}\sum_{j=1}^{N}{\bf \Phi}_{kj}$, $i=0,1,\cdots$. It can be validated that ${\bf W}_{i+1}={\bf W}_i+\sum_{k=1}^{N}{\bf \Phi}_{i+1,k}$, $i=0,1,\cdots$. Therefore, ${\bf W}_i\subseteq {\bf W}_{i+1}\subseteq \cdots \subseteq {\bf W}_{\infty}$, and ${\bf W}_{\infty}={\bar {\bf \Omega}}$. Note ${\bf W}_i$ is the sum of all possible items $A_N^{j_N}A_{N-1}^{j_{N-1}}\cdots A_{k}^{j_k}{\rm Im}\ B_k$ with $j_N+j_{N-1}+\cdots+j_k=i$, $j_N,...,j_k=0,1,...$, $k=1,...,N$. By the Cayley-Hamilton theorem, for any $j\ge l_0\doteq N(n-1)$, ${\bf W}_j\subseteq {\bf W}_{l_0}$. This means, ${\bf W}_{l_0}={\bf W}_{\infty}=\bar {\bf \Omega}$.
		

		Associated with $\{{\bf W}_i\}_{i=0}^{\infty}$, we construct the following matrix series
		\begin{equation}\label{matrix_series_cons}
		\begin{array}{c}
		{\Gamma}_{01}=B_1,{\Gamma}_{02}=B_2,...,{\Gamma}_{0N}=B_N\\
		\Gamma_{i1}=A_1\Gamma_{i-1,1},\Gamma_{i2}=A_2[\Gamma_{i-1,1},\Gamma_{i-1,2}],...,\\\Gamma_{iN}=A_N[\Gamma_{i-1,1},...,\Gamma_{i-1,N}],
		i=1,2,...,l_0.
		\end{array}
		\end{equation} 
		Let ${\cal W}_k={\bf row}\{\Gamma_{ij}|_{i=0,...,k}^{j=1,...,N}\}$. The above analysis indicates that ${\rm rank} {\cal W}_{l_0}= {\rm dim}  \bar {\bf \Omega}$.  Since this holds for all realizations of $(A_i,B_i)|_{i=1}^N$, we have ${\rm grank} {\cal W}_{l_0}= {\rm gdim}  \bar {\bf \Omega}$.
		
		Construct the dynamic graph
		${\tilde {\cal G}}\!=\!(\tilde V, \tilde E)$ associated with ${\cal W}_{{l_0}}$ as follows. The vertex set
		$\tilde V=\bigcup_{i=0}^{l_0} \tilde V_{i}$, with $\tilde V_i=\tilde V_{Ui}\cup \tilde V_{Xi}$, where
		$\tilde V_{Xi}=\{v_{ji}^k: j=1,...,n,k=1,...,N\}$, $\tilde V_{Ui}=\{v^{j}_{n+k,ti}:j=1,...,N,t=1,...,j,k=1,...,m_t\}$, for $i=1,...,{l_0}$, and $\tilde V_{X0}=\{v_{10},v_{20},...,v_{n0}\}$, $\tilde V_{U0}=\{v^j_{n+k,0}:j=1,...,N,k=1,...,m_j\}$. Let $\tilde V_U=\bigcup_{i=0}^{l_0}\tilde V_{Ui}$.
		We refer to the subgraph of $\tilde {\cal G}$ induced by $\tilde V_{i}$ as the $i$th layer, where the edge set $\tilde E_i$ within this layer is $\tilde E_i=\{(v^{p}_{n+k,ti},v^p_{qi}): p=1,...,N, t=1,...,p, k=1...,m_t, B_{t,qk}\ne 0\}$. We set $v^q_{p,0}\equiv v_{p,0}\in \tilde V_{X0}$, $\forall q$, and multiple edges are disallowed. As can be seen, in each layer except for the $0$th layer, $\tilde V_{Xi}$ consists of $N$ copies of state vertices, and $\tilde V_{Ui}$ consists of $N$ copies of input vertices, in which the $j$th copy consists of input vertices of the $j,j-1,...,1$th subsystems, $j=1,...,N$.   The edge set between the $i$ and $(i-1)$th layer $\tilde E_{i,i-1}$, $i=1,...,{l_0}$, is
		$$\tilde E_{i,i-1}\!=\!\{(v_{j,i}^k, v_{p,i-1}^q): k=1,...,N, q=k,...,N, A_{k,pj}\ne 0\}.$$
		The edge set of $\tilde {\cal G}$ is $\tilde E\!=\!(\bigcup_{i=0}^{l_0} \tilde E_i)\cup (\bigcup_{i=1}^{l_0}\tilde E_{i,i-1})$.  Since $\tilde {\cal G}$ has multiple layers, we call it the {\emph{multi-layer dynamic graph}} (MDG). Fig. \ref{dynamic-graph-temporal-multiple} illustrates $\tilde {\cal G}$ of a temporal network with $n=4$, $m_1=m_2=1$, and $N=2$. Compared to the CDG, the essential feature of an MDG is that, between two neighboring layers, there are edges from the $j$th copy of state vertices of the $i$th layer to the $j,j+1,...,N$th copies of state vertices of the $(i-1)$th layer, $j=1,...,N$, $i=1,...,l_0$.

		\begin{figure}
			\centering
			\includegraphics[width=3.3in]{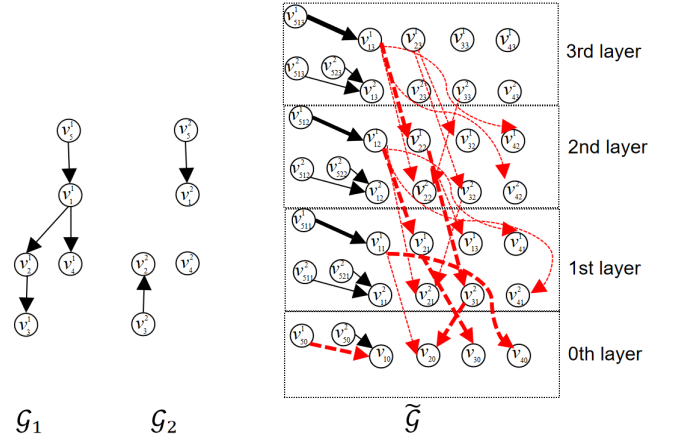}\\
			\caption{A temporal network and its associated MDG $\tilde {\cal G}$. Only the first $4$ layers of $\tilde {\cal G}$ are presented.  Dotted red lines represent edges between successive layers, and bold ones form a $\tilde V_U-\tilde V_{X0}$ linking of the maximum size $4$. }\label{dynamic-graph-temporal-multiple}
		\end{figure}
		
		
		The relation between $\tilde {\cal G}$ and ${\cal W}_{l_0}$ is explained as follows. Consider the $(t_k,t_0')$th entry of $A_{i_k}A_{i_{k-1}}\cdots A_{i_1}B_{i_0}$, which is a sub-column item in ${\cal W}_{l_0}$, $N\ge i_k\ge i_{k-1}\ge i_1\ge i_0$, $1\le t_k\le n$, $1\le t_0'\le m_{i_0}$. Such an entry is the sum of all the following items
		{\small{	\begin{equation}\label{nonzero-entry}A_{i_k}(t_k,t_{k-1})A_{i_{k-1}}(t_{k-1},t_{k-2})\cdots A_{i_1}(t_1,t_0)B_{i_0}(t_0,t_0'),\end{equation}}}where $t_k,t_{k-1},...,t_0=1,...,n$, and each involved entry is nonzero. By the construction of $\tilde {\cal G}$, there exists a {\emph{unique}} path
		{\small{$$P\doteq (v_{n+t_0',t_0k}^{i_0},v^{i_1}_{t_0,k},v_{t_1,k-1}^{i_2},\cdots, v_{t_{k-1},1}^{i_{k}},v_{t_k,0})$$}}with head in $\tilde V_{X0}$ and tail in $\tilde V_U$ in $\tilde {\cal G}$. The reverse direction also holds, that is, any $\tilde V_U-\tilde V_{X0}$ path, say $P$, corresponds to a {\emph{unique}} item in the form of (\ref{nonzero-entry}). Based on this relation,  we have the following result, whose proof is similar to that of Theorem \ref{upper-bound}, and therefore omitted. 
		
		%
		
		\begin{proposition}\label{necessary-reachable-set}
			The generic dimension of $\bar {\bf \Omega}$ is no more than the maximum size of a $\tilde V_U-\tilde V_{X0}$ linking in $\tilde {\cal G}$.
		\end{proposition}
		
		Since each layer of $\tilde {\cal G}$ contains $nN+\sum_{k=1}^{N}\sum_{i=1}^{k}m_i<nN^2$ vertices (assuming $m_i\le n$), $\tilde {\cal G}$ has no more than $nN^2l_0<N^3n^2$ vertices, which is only a small fraction of the original number, $n^N$, of items in ${\cal R}$. For the same reasoning in Remark \ref{reason-for-tight}, only under very strict conditions, the upper bound in Proposition \ref{necessary-reachable-set} will fail to match ${\rm gdim} \bar {\bf \Omega}$ exactly. In practice, Proposition \ref{necessary-reachable-set} can provide a good estimation of ${\rm gdim} \bar {\bf \Omega}$. For example, the MDG in Fig. \ref{dynamic-graph-temporal-multiple} has a maximum size $4$ of linkings, which is consistent with ${\rm grank}{\cal R}$ obtained by assigning random values to the corresponding nonzero entries. In the following, we give an interesting corollary of this proposition.
		
		Recall ${\cal G}_i$ is the system digraph of the $i$th subsystem. Define the {\emph{set of switching edges}} $E_{\rm {sw}}=\{(v_j^i,v_j^k): i=1,...,N-1,k=i+1,...,N,j=1,...,n\}$, representing the switching direction of subsystems.
		The digraph ${\cal G}_{\rm {sw}}$ is defined as the graph $\bigcup_{i=1}^N {\cal G}_i$ with the addition of switching edge set $E_{\rm {sw}}$, which can depict the signal flow over the state and input vertices of subsystems.
		\begin{definition}
			A state vertex $v\in V_{A_i}$ is said to be input-reachable in ${\cal G}_{\rm {sw}}$, if there is a path starting from an input vertex $u\in \bigcup_{i=1}^N V_{B_i}$ ending at $v$ in ${\cal G}_{\rm {sw}}$.
		\end{definition}
		
		Let $\bar {\cal G}_{\rm {sw}}$ be the digraph obtained from ${\cal G}_{\rm {sw}}$ after deleting all input-unreachable vertices in ${\cal G}_{\rm {sw}}$ (and their incident edges). Let $A_1',...,A_N'$ be the resulting matrices of $A_1,...,A_N$ by replacing the rows and columns corresponding to the input-unreachable vertices with zeros.
		The following result gives a necessary condition for the full dimensionality of $\bar {\bf \Omega}$ (so for the structural overall reachability). 
		\begin{corollary} \label{necessary-dynamic-linking}
			A necessary condition for ${\rm gdim} \bar {\bf \Omega}= n$ is that, ${\rm grank}[A_1',,...,A_N',B_1,...,B_N]=n$.
		\end{corollary}
		\begin{proof}
			By Proposition \ref{necessary-reachable-set}, it is necessary for the MDG $\tilde{\cal G}$ to have a $\tilde V_U-\tilde V_{X0}$ linking with size $n$. Denote such a linking by $L$.
			By the construction of $\tilde {\cal G}$, all head vertices $\tilde V_{X0}$ of $L$,  should be input-reachable. Moreover, there exist $n$ disjoint edges between $\tilde V_{U0}\cup \tilde V_{X1}$ and $\tilde V_{X0}$. These conditions are equivalent to that, a matching with size $n$ exists in the bipartite graph $(\tilde V_{U0}\cup \tilde V_{X1}\backslash \tilde V_{\rm unr}, \tilde V_{X0}, \tilde E)$, where $\tilde V_{\rm unr}$ is the subset of $\tilde V_{X1}$ that cannot be the heads of paths from the input vertices in $\tilde {\cal G}$, and $\tilde E\subseteq \tilde E_{10}\cup \tilde E_{0}$ are the edges between the corresponding bipartitions. The corollary then follows immediately from the equivalence between the generic rank of a structured matrix and the maximum matching of its associated bipartite graph \citep[Proposition 2.1.12]{Murota_Book}. 
		\end{proof}
		
		\begin{remark} Interestingly, the condition in Corollary \ref{necessary-dynamic-linking} is claimed to be necessary and sufficient for the structural controllability of switched systems in \cite{LiuStructural}. However, by virtue of Corollary  \ref{hard-minimal-space}, such a condition cannot be sufficient for the full dimensionality of $\bar {\bf \Omega}$ for  temporal networks.
			
		\end{remark}

		\subsection{Lower bound of the dimension of $\bar {\bf \Omega}$} \label{sec-low}
		This subsection gives a graph-theoretic lower bound of ${\rm gdim}{\bar {\bf \Omega}}$ based on the so-called temporal cactus.

		\begin{definition}[Temporal cactus] \label{temporal-cactus} A temporal stem is a path with no repeated vertices from an input vertex in $\bar {\cal G}_{\rm {sw}}$. A temporal cactus configuration is a collection of vertex-disjoint temporal stems and cycles in $\bar {\cal G}_{\rm {sw}}$.
		\end{definition}
		\begin{definition}[Temporal cactus covering]
			A vertex $v_j^i\in V_{A_i}$ is said to be covered by a temporal cactus configuration ${\cal F}$ if it is the head vertex of an edge {\bf{but the switching edge}} of ${\cal F}$.
			A set $V\subseteq X\doteq \{v_1,...,v_n\}$ is said to be covered by a temporal cactus configuration, if there is a temporal cactus ${\cal F}$ so that for each $v_j\in V$, there exists some $i\in \{1,...,N\}$ such that $v_j^i$ is covered by ${\cal F}$.
		\end{definition}

		
		
		\begin{example}\label{exp3} To illustrate the temporal cactus configuration, Fig. \ref{general-cactus} shows the $\bar {\cal G}_{\rm {sw}}$ of a temporal network with $n=4$, $N=2$. The subgraph consisting of a temporal stem $(v_5^1,v_1^1,v_1^2, v_4^2)$ and a cycle $(v_3^2,v_3^2)$ forms a temporal cactus configuration that covers $\{v_1,v_4,v_3\}$. The following Theorem \ref{lower-bound} yields ${\rm gdim}{\bar {\bf \Omega}}\ge 3$.
			\begin{figure}
				\centering
				\includegraphics[width=2.92in]{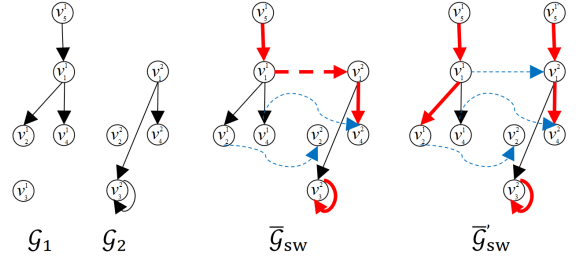}\\
				\caption{A temporal network with its subsystem digraphs being ${\cal G}_1$ and ${\cal G}_2$, and the associated ${\bar {\cal G}}_{\rm {sw}}$ (note $v_3^1$ is not input-reachable). Dotted blue lines represent the switching edges. The bold red lines constitute a temporal cactus configuration that covers $\{v_1,v_4,v_3\}$. See Example \ref{second-example} for the explanation of ${\bar {\cal G}}'_{\rm {sw}}$.}\label{general-cactus}
			\end{figure}
		\end{example}
		
		\begin{theorem}\label{lower-bound}
			The generic dimension of $\bar {\bf \Omega}$ is no less than the size of any set $V\subseteq X$ that can be covered by a temporal cactus configuration in $\bar {\cal G}_{\rm {sw}}$.
		\end{theorem}
		
		\begin{proof} Let $V=\{v_{i_1},...,v_{i_p}\}$ be covered by a temporal cactus configuration ${\cal F}$. Then, there are vertices $v^{j_1}_{i_1},...,v^{j_p}_{i_p}$ that are covered by ${\cal F}$ in $\bar {\cal G}_{\rm {sw}}$, $1\le j_1,...,j_p\le N$. It is claimed that there is a linking of size $p$ from some $\tilde V_U'\subseteq \tilde V_U$ to $\{v_{i_10},...,v_{i_p0}\}$ in $\tilde {\cal G}$ ($|\tilde V_U'|=p$), denoted by $L$.
			A simple way to show this is to regard ${\cal G}_{\rm {sw}}$ as the system digraph of a single system ${\bf \Sigma }$ with $nN$ state vertices $\{v_1^1,...,v_n^1,...,v_1^N,v_n^N\}$. Then, by \cite{poljak1990generic}, all vertices covered by $\cal F$ in $\bar {\cal G}_{\rm {sw}}$ induces a linking of an appropriate size in the corresponding dynamic graph of ${\bf \Sigma }$, in which vertices that are heads of the switching edges are not included.  Such a linking corresponds to a linking with size $p$ in $\tilde  {\cal G}$ by the construction.
			
			By assigning unit weights to the switching edges, construct the multigraph $H_L$ associated with the linking $L$, which consists of vertices and edges of ${\cal F}$, and the corresponding paths from input vertices to one vertex of each cycle (there is a one-to-one correspondence between edges of $H_L$ and $L$; see Remark \ref{reason-for-tight} for the definition of $H_L$). Since the temporal stems and cycles are vertex-disjoint, it can be seen that no other $\tilde V_U'-\{v_{i_10},...,v_{i_p0}\}$ linking in $\tilde {\cal G}$ can produce the weight equaling to $w(L)$, as any other $\tilde V_U'-\{v_{i_10},...,v_{i_p0}\}$ linking must introduce some new edges or reduce some old edges of $H_L$ due to its vertex-disjointness. By the similar reasoning in (\ref{det-important}), we know ${\rm gdim}{\bar {\bf \Omega}}$ is at least $p$.
		\end{proof}
		
		
		Obviously, for a single system, the temporal cactus configuration reduces to the conventional one introduced in \cite{Lin_1974}, which exactly characterizes the generic dimension of the controllable subspace \cite{hosoe1980determination}. The following proposition indicates, when applied to temporal networks with dedicated inputs, i.e., each input actuates at most one state per subsystem, which is often the case for large-scale networks with geographically distributed nodes {\cite{P.Fa2014Controllability}}, Theorem  \ref{lower-bound} may be more effective to provide the lower bounds.
		
		\begin{proposition} \label{dedicated-networks}
			Consider two temporal networks ${\bf \Sigma}_1$ and ${\bf \Sigma}_2$ with structured parameters $(A_i,B_i)|_{i=1}^N$ and $(A_i,B'_i)|_{i=1}^N$, in which
			$B_1=I^{0}_{S_1},B_2=I^{0}_{S_2},...,B_N=I^{0}_{S_N}$, and $B_1'=I^{1}_{S_1}, B_2'=[I^{2}_{S_1}, I^{2}_{S_2}],...,B_N'=[I^{N}_{S_1},...,I^{N}_{S_N}]$,
			where $S_1,...,S_N$ $\subseteq \{1,...,n\}$, $I^{j}_{S_i}$ is structurally equivalent to the sub-columns of $I_n$ indexed by $S_i$, and for $k\ne j$, $I^{j}_{S_i}$ and $I^{k}_{S_i}$ are independent, $k,j=0,...,N$. The subspaces $\bar {\bf \Omega}$ associated with ${\bf \Sigma}_1$ and ${\bf \Sigma}_2$ are denoted by $\bar {\bf \Omega}_{{\bf \Sigma}_1}$ and $\bar {\bf \Omega}_{{\bf \Sigma}_2}$.
			Then, ${\rm gdim} \bar {\bf \Omega}_{{\bf \Sigma}_1}= {\rm gdim} \bar {\bf \Omega}_{{\bf \Sigma}_2}.$
		\end{proposition}
		
		
		\begin{proof}
			Consider the subspaces $\{{\bf \Phi}_{ij}\}$, $\{{\bf W}_i\}$, and $\{{\bf \Phi}'_{ij}\}$, $\{{\bf W}'_i\}$ associated with ${\bf \Sigma}_1$ and ${\bf \Sigma}_2$ respectively, whose constructions are given in (\ref{subspaces_series_cons}). For almost all realizations of $\{I_{S_i}^j\}_{i=1,...N}^{j=0,...,N}$, we have
			$$ \begin{aligned} {\bf \Phi}_{11}&=A_1{\rm Im} I^0_{S_1}=A_1{\rm Im} I^1_{S_1}={\bf \Phi}'_{11}, \\
			{\bf \Phi}_{12}&=A_2{\rm Im}[I^0_{S_1},I^0_{S_2}]=A_2{\rm Im} [I^1_{S_1},I^2_{S_1}, I^2_{S_2}]={\bf \Phi}'_{12}, \\
			&\vdots\\
			{\bf \Phi}_{1N}&=A_N{\rm Im}[I^0_{S_1},...,I^0_{S_N}]\\&=A_N{\rm Im} [I^1_{S_1},I^2_{S_1}, I^2_{S_2},...,I^N_{S_1},...,I^N_{S_N}]={\bf \Phi}'_{1N}
			\end{aligned}$$
			By the constructions of $\{{\bf \Phi}_{ij}\}$ and $\{{\bf \Phi}'_{ij}\}$, we have ${\bf \Phi}_{ij}={\bf \Phi}'_{ij}$ for $i\ge 1$, $j=1,...,N$. Moreover, ${\bf W}_0=\sum_{i=1}^N {\rm Im} B_{i}=\sum_{i=1}^N{\rm Im}B'_{i}={\bf W}'_0$. This leads to ${\bf W}_{l_0}={\bf W}'_{l_0}$. The proposed result then follows immediately. 
		\end{proof}
		
		Proposition \ref{dedicated-networks} roughly says, for temporal networks\\ $(A_i,B_i)|_{i=1}^N$ {\emph{with dedicated inputs}}, we can replace the input matrix $B_k$ with $[B_1,...,B_{k-1},B_k]$, $k=2,...,N$, such that the {\emph{generic dimension}} of the resulting $\bar {\bf \Omega}$ remains unchanged. This may increase the number of temporal stems. 

		\begin{example}[Example \ref{exp3} cont] \label{second-example}
			Consider the temporal network in Fig. \ref{general-cactus}. Theorem \ref{lower-bound} on the original system yields a lower bound $3$ for ${\rm gdim} {\bar {\bf \Omega}}$. However, Theorem \ref{lower-bound} on the equivalent system (with a new input $v_5^{1'}$ on $v_1^2$) yields a lower bound $4$ (see the bold red lines of $\bar {\cal G}'_{\rm {sw}}$ in Fig. \ref{general-cactus}), indicating the full dimensionality of $\bar {\bf \Omega}$. 
		\end{example}
		
		{\bf Computation issues:} Finding the {\emph{maximum size}} of a set $V^*\subseteq X$ that can be covered by a temporal cactus configuration in $\bar {\cal G}_{\rm {sw}}$ can give a good lower bound of ${\rm gdim}{\bar {\bf \Omega}}$. However, since typically a $V$ can correspond to an exponential number (in $N$) of combinations of subsystem vertices, finding a maxima $V^*$ is not easy. We provide the following heuristic algorithm to get a lower bound of $V^*$ in polynomial time (let the bipartite graph ${\cal B}_{\rm sw}\doteq {\cal B}([A'_1,...,A'_N,B_1,...,B_N])$):
		\begin{itemize}
			\item[(1)] Determine a maximum matching ${\cal M}$ in ${\cal B}_{\rm sw}$, and map it
			to a vertex set $W$: the $(i,j)$th entry of $A'_k$ ($B_k$) associated to an edge of ${\cal M}$ is mapped to the vertex $v^k_i$, constituting a member of $W$;
			\item[(2)] Find a maximum subset $V'\subseteq W$ which can be covered by a cactus configuration in $\bar {\cal G}_{\rm {sw}}$ in the sense of Definition \ref{cactus-original}, using the maximum weighted matching algorithm \cite{murota1990note} (see Step 3 of Algorithm \ref{alg1}).
		\end{itemize}We can implement the above computation for various ${\cal M}$ and take the maxima as a lower bound of ${\rm gdim}{\bar {\bf \Omega}}$.  The reason why we choose the matching ${\cal M}$ is that, if $V\subseteq X$ is covered by a temporal cactus configuration, there must be some $v^{j}_i\in V_{A_{j}}$ per $v_i\in V$, such that $\{v^{j}_i: i\in V\}$ are covered by a matching in ${\cal B}_{\rm sw}$.

		\section{Applications to structured switched systems} \label{sec-application}
		We briefly show the applications of the previous results to structured switched systems. The controllability/reachability of switched systems has been completely characterized in \cite{Z.S2002Controllability} as follows.
		
		\begin{lemma}\cite{Z.S2002Controllability} \label{switched-criteria}
			The switched system (\ref{plant-switched}) is controllable (reachable), if and only if the following matrix ${\cal R}_{\rm sw}$ has full row rank:
			$${\cal R}_{\rm sw}={\bf row}\left\{ A_{i_N}^{j_N}A_{i_{N-1}}^{j_{N-1}}\cdots A_{i_1}^{j_1} B_{i_1}|_{i_1,...,i_N=1,...,N}^{j_N,...,j_k=0,...,n-1}\right\}. $$
			Moreover, the dimension of the controllable (reachable) subspace of system (\ref{plant-switched}) is ${\rm rank} {\cal R}_{\rm sw}$.
		\end{lemma}	
		{\bf{Application to controllability realization:}}
		We first show an application of the upper bound of ${\rm gdim}{\bf \Omega}_{\{h_i\}}$ in the {\emph{controllability realization problem (CRP)}} of switched systems. Given a switching path $\sigma=(i_m)_{m=1}^l$, where $i_m\in\{1,...,N\}$ and $l\le \infty$ is the length of $\sigma$, a natural question is that, is it possible to design $\{h_m\}_{m=1}^l$, where $h_m$ is the time duration of the $m$th subsystem realization, $m\in\{1,...,l\}$, so that the reachable subspace of system (\ref{plant-switched}) is ${\mathbb R}^n$?  Such a problem, known as (a variant) the CRP, has been studied in much literature in the context of numerical systems \cite{Z.S2002Controllability,xie2003controllability}. However, a similar problem has not yet been studied in the generic sense, i.e., in the structured system theory.
		
		A similar argument to Lemmas \ref{generic-dimension} and \ref{generic-dimension-2} can show that, given the switching path $\sigma$, the maximum dimension of the associated reachable subspaces (over all $\{h_m\}_{m=1}^l$) is a generic property for $\phi\in {\mathbb R}^d$. Combined with these facts, our method in Theorem \ref{upper-bound} can provide a necessary condition for the following problem:
		{\emph{Given the structured switched system (\ref{plant-switched}) and a switching path $\sigma=(i_m)_{m=1}^l$, is there generically a series of time duration $\{h_m\}_{m=1}^l$, such that the associated reachable subspace is ${\mathbb R}^n$?}}
		
		To this end, define $${\cal C}'=[C_{i_l},e^{A_{i_l}h_l}C_{i_{l-1}},\cdots, e^{A_{i_l}h_l}\cdots e^{A_{i_2}h_2}C_{i_1}],$$so that ${\rm span}{\cal C}'$ is the reachable subspace of the switched system with the switching path $(i_m)_{m=1}^l$ and the time durations $\{h_m\}_{m=1}^l$ \cite{Z.S2002Controllability,xie2003controllability}. Let $(A'_{m},B'_{m})$ be structurally equivalent to $(A_{i_m},B_{i_m})$, $m\in\{1,...,l\}$. Introduce ${\cal C}'$ associated to the temporal system $(A'_{m},B'_{m})|_{m=1}^l$ in the same way as (\ref{key-surrogate}):
		\begin{equation}
		\label{switch-subspace}\bar {\cal C}'=[C'_{l}, M'_{l}C'_{{l-1}},\cdots, M'_{l}M'_{{l-1}}\cdots M'_{2}C'_{1}].\end{equation}
		It is easy to see ${\rm grank}\bar {\cal C}'\ge {\rm grank} {\cal C}'$.  Construct the CDG ${\bar {\cal G}}'$ associated with $\bar {\cal C}'$. Then, a necessary condition for the above problem is that, the maximum size of a $\bar V_B-\bar V_{A_l,n}$ linking in $\bar {\cal G}'$ should be $n$. For the same reasoning in Remark \ref{reason-for-tight}, we believe this could also provide the sufficient condition for a large class of switched systems.  The following example is one such.
		
		\begin{example}\label{switched-example} Consider a switched system with $2$ subsystems, whose digraphs ${\cal G}_1$ and ${\cal G}_2$ are given in Fig. \ref{first-example}. The associated CDGs' with the switching paths $\sigma_1=(1,2)$ and $\sigma_2=(1,2,1)$, denoted by  $\bar {\cal G}_{\sigma_1}$ and $\bar {\cal G}_{\sigma_2}$, are given in Fig. \ref{switch-design}. From them, we know the switching path $\sigma_1$ is not sufficient for a full dimensional reachable subspace, while $\sigma_2$ satisfies the necessary condition (as the maximum size of the corresponding linking is $3$). By assigning the number $1$ to all nonzero entries of $(A_1,B_1)$ and $(A_2,B_2)$, it can be computed that the reachable subspace with $\sigma_1$ is
			${\rm span }\ e^{A_2h_2}C_1={\rm span}{\footnotesize{\left[\begin{array}{cc}
					1 & 0\\
					0 & 1\\
					0 & h_2\\
					\end{array}\right]}}$, and with $\sigma_2$ is
			${\rm span }[C_1,e^{A_1h_3}e^{A_2h_2}C_1]={\rm span}{\footnotesize{\left[\begin{array}{cccc}
					1 & 0 & 1 & 0\\
					0 & 1 & h_3 & 0\\
					0 & 0 & 0 & h_2\\
					\end{array}\right]}}$, for any fixed $h_2,h_3>0$. Hence, only $\sigma_2$ leads to a full dimensional reachable subspace.
			\begin{figure}
				\centering
				\includegraphics[width=2.3in]{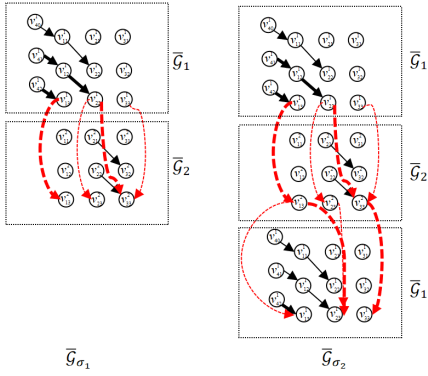}\\
				\caption{The associated CDGs' to a switched system with subsystem digraphs ${\cal G}_1$ and ${\cal G}_2$ given in Fig. \ref{first-example}, under two switching paths $\sigma_1=(1,2)$ and $\sigma_2=(1,2,1)$. Bold lines form a linking with the maximum size in the respective graphs. }\label{switch-design}
			\end{figure}
		\end{example}
		
		
		The above example motivates a natural problem, that is, given the structured switched system (\ref{plant-switched}), how to design a switching path $\sigma$ with the smallest length $l$ so that the associated reachable subspace is ${\mathbb R}^n$. Such a problem has been extensively studied for numerical systems; see \cite{ji2008controllability}.
		If we can determine the smallest $l^*$ {\emph{combinatorially}} so that $\bar {\cal C}'$ has full row generic rank, then $l^*$ can provide a {\emph{lower bound}} for the aforementioned problem. 
		
		{\bf Application to the generic dimension of controllable subspace:} Although some characterizations for structural controllability of switched systems have been given in \cite{LiuStructural}, they shed no light on the controllable subspace if the system is not structurally controllable. We briefly discuss the application of the lower bound of ${\rm gdim}{\bar {\mathbf{\Omega}}}$ in estimating the generic dimension of controllable subspace of switched systems.
		
		By comparing ${\cal R}_{\rm sw}$ in Lemma \ref{switched-criteria} and ${\cal R}$ in (\ref{R_def}), it is obvious ${\rm grank} {\cal R}_{\rm sw}\ge {\rm grank} {\cal R}$. If we change the switching path $(1,2,...,N)$ to $\sigma$, which is a permutation of $(1,2,...,N)$, then the generic dimension of $\bar {\bf \Omega}$ of the resulting temporal network, denoted by ${\rm gdim} {\bar {\bf \Omega}}_{\sigma}$, is a {\emph{lower bound}} of ${\rm grank} {\cal R}_{\rm sw}$. Hence, based on the procedure in Section \ref{sec-low}, we can run across various permutations $\sigma$'s and take the maximum ${\rm gdim} {\bar {\bf \Omega}}_{\sigma}$, which is a lower bound of ${\rm grank} {\cal R}_{\rm sw}$.
		
		\begin{example}
			Consider a structured switched system with two subsystems, whose system digraphs are ${\cal G}_1$ and ${\cal G}_2$ in Fig. \ref{switch-design}. From the analysis in Example \ref{second-example}, we know for the corresponding temporal system, ${\rm grank}{\cal R}=4$, leading to ${\rm grank} {\cal R}_{\rm sw}=4$ (as $n=4$). Indeed, one can check that $[B_1,A_1B_1,A_2B_1,A_2^2B_1]$ is generically non-singular, indicating that this switched system is structurally controllable. 
		\end{example}
		
		\section{Conclusion}
		In this paper, the reachability and controllability of temporal {\emph{continuous-time}} networks are analyzed from a generic standpoint. It is proven that the dimension of the reachable subspace on a single temporal sequence is generic w.r.t. the parameters of subsystem matrices and the time duration of each subsystem. Explicit expressions of the minimal subspace containing the overall reachable set are given. Unlike in the case of temporal discrete-time networks, finding verifiable conditions for structural (overall)  reachability of continuous-time networks is shown to be hard.  Graph-theoretic lower and upper bounds are given for the generic dimension of ${\mathbf{\Omega}}_{\{h_i\}}$ and $\bar {\mathbf{\Omega}}$, which generalize the classical dynamic graph and cactus notions to temporal networks. These results can be applied to the analysis of structured switched systems. All results can be directly extended to their controllability counterparts by reversing the order of subsystems.
		
		
		The study raises some interesting questions, such as the possibility of the method in Section \ref{sec-application} of providing an upper bound for the length of a switching path in the CRP, an open issue in switched system theory \cite{ji2008controllability}, and the determination of the exact generic dimension of controllable subspaces of structured switched systems. To address these questions, the cactus notion may need to be further generalized for switched systems.
		
		\section*{Acknowledgments}
		The authors are grateful to Prof. Zhendong Sun of Shandong University of Science \& Technology for valuable discussions and suggestions on the manuscript of this work. 
		
		\section*{Appendix}
		{\bf Proof of Proposition \ref{boud-alg1}}: Let ${\cal I}_0=\{1,...,N\}$, and $\tilde V^*=V_1^*\cup \cdots \cup V_N^*$ be an optimal solution to problem $\cal P$. Suppose Algorithm \ref{alg1} selects $V_{i^*_1}$, $V_{i^*_2}$,...,$V_{i^*_k}$ successively during the iterations, with $k\le N$, and $\tilde V_{j}=V_{i_1^*}\cup\cdots\cup V_{i_j^*}$, $1\le j\le k$.\footnote{For the ease of notation, we define $\tilde V_{j}=\tilde V_k$ for $j=k+1,...,N$.} Consider the first iteration. Since $|\tilde V^*|\le \sum_{i=1}^N|V_i^*|$, we have $\max |V_i^*|\doteq \max_{i\in {\cal I}_0}|V_i^*|\ge |\tilde V^*|/N$. As $\tilde V^* \subseteq E_0=X$, by definition, $|V_{i_1^*}|\ge \max |V_i^*|\ge |\tilde V^*|/N$.
		Consider the second iteration. Observe that $|V_{i_1^*}\bigcup (\bigcup \nolimits_{i\in {\cal I}_0\backslash \{i_1^*\}} V_i^*)|\ge |\bigcup \nolimits_{i\in {\cal I}_0\backslash \{i_1^*\}}  V_i^*|\ge |\tilde V^*|-|V^*_{i_1^*}|\ge |\tilde V^*|-|V_{i_1^*}|$. Therefore, there must be an $i\in {\cal I}_0\backslash \{i_1^*\}$, such that $|V_{i_1^*}\cup V_i^*|-|V_{i_1^*}|\ge (|\tilde V^*|-2|V_{i_1^*}|)/(N-1)$; otherwise $|V_{i_1^*}\bigcup (\bigcup \nolimits_{i\in {\cal I}_0\backslash \{i_1^*\}} V_i^*)|< |V_{i^*_1}|+(N-1)*(|\tilde V^*|-|V_{i_1^*}|-|V_{i_1^*}|)/(N-1)=|\tilde V^*|-|V_{i_1^*}|$. By the definition of $V_{i_2^*}$,
		$|V_{i^*_1}\cup V_{i^*_2}|-|V_{i^*_1}|\ge (|\tilde V^*|-2|V_{i_1^*}|)/(N-1)$.
		Consequently, for $N\ge 2$,  $$|\tilde V_2|\ge |V_{i^*_1}|+\frac{|\tilde V^*|-2|V_{i_1^*}|}{N-1}=\frac{|\tilde V^*|}{N-1}+\frac{N-3}{N-1}|V_{i_1^*}|.$$ By substituting $|V_{i_1^*}|\ge |\tilde V^*|/N$ into the above formula, we get that, for $N\ge 3$,
		$|\tilde V_k|\ge |\tilde V_2|\ge \frac{2N-3}{N(N-1)}|\tilde V^*|,$ which is (\ref{relation1}).
		As a result, if $N=2$, $|\tilde V_k|\ge |V_{i^*_1}|\ge \frac{1}{2}|\tilde V^*|$. If $N=3$, $|\tilde V_k|\ge \frac{1}{2}|\tilde V^*|$. This proves the approximation factor $\frac{1}{2}$ for $N=2,3$.
		
		Let us consider the $(t+1)$th iteration ($t\ge 2$). Similar to the above, we observe $|\tilde V_{t}\bigcup (\bigcup \nolimits_{i\in {\cal I}_0\backslash \{i_1^*,...,i_t^*\}}V_i^*)|\ge |\bigcup \nolimits_{i\in {\cal I}_0\backslash \{i_1^*,...,i_t^*\}}V_i^* |\ge |\tilde V^*|-|\bigcup \nolimits_{i\in \{i_1^*,...,i_t^*\}}V_i^*|\ge |\tilde V^*|-t|V_{i_1^*}|$, where $|\bigcup \nolimits_{i\in \{i_1^*,...,i_t^*\}}V_i^*|\le \sum \nolimits_{j=1}^t |V^*_{i_j^*}|\le t\cdot\max\nolimits |V_i^*|\le t|V_{i_1^*}|$ has been used. Hence, there is an $i\in {\cal I}_0\backslash \{i_1^*,...,i_t^*\}$ satisfying $|\tilde V_{t}\cup V_{i}^*|-|\tilde V_{t}|\ge (|\tilde V^*|-t|V_{i^*_1}|-|\tilde V_t|)/(N-t)$; otherwise $|\tilde V_{t}\bigcup (\bigcup \nolimits_{i\in {\cal I}_0\backslash \{i_1^*,...,i_t^*\}}V_i^*)|\le |\tilde V_t|+\sum \limits_{j\in {\cal I}_0\backslash \{i_1^*,...,i_t^*\}} (|\tilde V_{t}\cup V_j^*|-|\tilde V_{t}|)<|\tilde V^*|-t|V_{i_1^*}|$. Since $V=V_{i^*_{t+1}}$ maximizes $|V\cup \tilde V_t|-|\tilde V_t|$ over $V\in U_{i}, i\in {\cal I}_0\backslash \{i_1^*,...,i_t^*\}$, we have $|\tilde V_{t}\cup V_{i_{t+1}^*}|-|V_{i_{t+1}^*}|\ge (|\tilde V^*|-t|V_{i^*_1}|-|\tilde V_t|)/(N-t)$. This leads to $|\tilde V_{t+1}|\ge |\tilde V_t|+ (|\tilde V^*|-t|V_{i^*_1}|-|\tilde V_t|)/(N-t)$, which yields
		\begin{equation} \label{iteration}
		\frac{|\tilde V^*|-|\tilde V_{t+1}|}{N-t-1}- \frac{|\tilde V^*|-|\tilde V_{t}|}{N-t}\le \frac{t}{(N-t-1)(N-t)}|V_{i_1^*}|,
		\end{equation}
		for $N\ge 3$ and $t<N-1$. Taking the summation of both sides of (\ref{iteration}) for $t$ from $1$ to $N-2$, we get
		$$\frac{|\tilde V^*|-|\tilde V_{t+1}|}{N-t-1}\le \frac{|\tilde V^*|-|V_{i_1^*}|}{N-1}+ \sum_{j=1}^t\frac{j}{(N-j-1)(N-j)}|V_{i_1^*}|,$$
		which leads to $|\tilde V_{t+1}|\ge f(t)$, for $1\le t\le N-2$. As $|\tilde V_k|\ge |\tilde V_{t}|$ for any $t=1,...,N$,
		we prove (\ref{relation2}).  
		
		\bibliographystyle{elsarticle-num}
		{\small
			\bibliography{yuanz3}

\begin{thebibliography}{10}
\expandafter\ifx\csname url\endcsname\relax
  \def\url#1{\texttt{#1}}\fi
\expandafter\ifx\csname urlprefix\endcsname\relax\def\urlprefix{URL }\fi
\expandafter\ifx\csname href\endcsname\relax
  \def\href#1#2{#2} \def\path#1{#1}\fi

\bibitem{posfai2014structural}
M.~P{\'o}sfai, P.~H{\"o}vel, Structural controllability of temporal networks,
  New Journal of Physics 16~(12) (2014) 123055.

\bibitem{li2017fundamental}
A.~Li, S.~P. Cornelius, Y.-Y. Liu, L.~Wang, A.-L. Barab{\'a}si, The fundamental
  advantages of temporal networks, Science 358~(6366) (2017) 1042--1046.

\bibitem{holme2005network}
P.~Holme, Network reachability of real-world contact sequences, Physical Review
  E 71~(4) (2005) 046119.

\bibitem{barabasi2005origin}
A.-L. Barabasi, The origin of bursts and heavy tails in human dynamics, Nature
  435~(7039) (2005) 207--211.

\bibitem{kossinets2008structure}
G.~Kossinets, J.~Kleinberg, D.~Watts, The structure of information pathways in
  a social communication network, in: Proceedings of the 14th ACM KDD, 2008,
  pp. 435--443.

\bibitem{wei2022distributed}
M.~Wei, Y.~Xia, Y.~Zhang, Y.~Zhan, B.~Cui, Distributed consensus control for
  networked euler--lagrange systems over directed graphs: A dynamic
  event-triggered approach, International Journal of Robust and Nonlinear
  Control 32~(16) (2022) 8786--8803.

\bibitem{Y.Y.2011Controllability}
Y.~Y. Liu, J.~J. Slotine, A.~L. Barabasi, Controllability of complex networks,
  Nature 48~(7346) (2011) 167--173.

\bibitem{Ramos2022AnOO}
G.~Ramos, A.~P. Aguiar, S.~D. Pequito, An overview of structural systems
  theory, Automatica 140 (2022) 110229.

\bibitem{zhang2019structural}
Y.~Zhang, T.~Zhou, Structural controllability of an {NDS} with {LFT}
  parameterized subsystems, IEEE Transactions on Automatic Control 64~(12)
  (2019) 4920--4935.

\bibitem{S.Pe2016A}
S.~Pequito, S.~Kar, A.~P. Aguiar, A framework for structural input/output and
  control configuration selection in large-scale systems, IEEE Transactions on
  Automatic Control 48~(2) (2016) 303--318.

\bibitem{zhang2019minimal}
Y.~Zhang, T.~Zhou, Minimal structural perturbations for controllability of a
  networked system: Complexities and approximations, International Journal of
  Robust and Nonlinear Control 29~(12) (2019) 4191--4208.

\bibitem{zhang2021unified}
Y.~Zhang, V.~Latora, A.~E. Motter, Unified treatment of synchronization
  patterns in generalized networks with higher-order, multilayer, and temporal
  interactions, Communications Physics 4~(1) (2021) 195.

\bibitem{conner1987structure}
L.~T. Conner~Jr, D.~P. Stanford, The structure of the controllable set for
  multimodal systems, Linear Algebra and its Applications 95 (1987) 171--180.

\bibitem{ezzine1989controllability}
J.~Ezzine, A.~Haddad, Controllability and observability of hybrid systems,
  International Journal of Control 49~(6) (1989) 2045--2055.

\bibitem{sun2001reachability}
Z.~Sun, D.~Zheng, On reachability and stabilization of switched linear systems,
  IEEE Transactions on Automatic Control 46~(2) (2001) 291--295.

\bibitem{Z.S2002Controllability}
Z.~Sun, S.~S. Ge, T.~H. Lee, Controllability and reachability criteria for
  switched linear systems, Automatica 48~(5) (2002) 775--786.

\bibitem{xie2003controllability}
G.~Xie, L.~Wang, Controllability and stabilizability of switched
  linear-systems, Systems \& Control Letters 48~(2) (2003) 135--155.

\bibitem{stanford1980controllability}
D.~P. Stanford, L.~T. Conner, Jr, Controllability and stabilizability in
  multi-pair systems, SIAM Journal on Control and Optimization 18~(5) (1980)
  488--497.

\bibitem{ge2001reachability}
S.~S. Ge, Z.~Sun, T.~H. Lee, Reachability and controllability of switched
  linear discrete-time systems, IEEE Transactions on Automatic Control 46~(9)
  (2001) 1437--1441.

\bibitem{LiuStructural}
X.~Liu, H.~Lin, B.~M. Chen, Structural controllability of switched linear
  systems, Automatica 49~(12) (2013) 3531--3537.

\bibitem{Lin_1974}
C.~T. Lin, Structural controllability, IEEE Transactions on Automatic Control
  19~(3) (1974) 201--208.

\bibitem{pequito2017structural}
S.~Pequito, G.~J. Pappas, Structural minimum controllability problem for
  switched linear continuous-time systems, Automatica 78 (2017) 216--222.

\bibitem{zhang2022constrained}
Y.~Zhang, Y.~Xia, S.~Liu, Z.~Su, On polynomially solvable constrained input
  selections for fixed and switched linear structured systems,
  arXiv:2204.01084, 2022.

\bibitem{poljak1990generic}
S.~Poljak, On the generic dimension of controllable subspaces, IEEE
  Transactions on Automatic Control 35~(3) (1990) 367--369.

\bibitem{zhang2021higher}
Y.~Zhang, A.~Garas, I.~Scholtes, Higher-order models capture changes in
  controllability of temporal networks, Journal of Physics: Complexity 2~(1)
  (2021) 015007.

\bibitem{hou2016structural}
B.~Hou, X.~Li, G.~Chen, Structural controllability of temporally switching
  networks, IEEE Transactions on Circuits and Systems I: Regular Papers 63~(10)
  (2016) 1771--1781.

\bibitem{nicosia2013graph}
V.~Nicosia, J.~Tang, C.~Mascolo, M.~Musolesi, G.~Russo, V.~Latora, Graph
  metrics for temporal networks, in: Temporal networks, Springer, 2013, pp.
  15--40.

\bibitem{antsaklis1997linear}
P.~J. Antsaklis, A.~N. Michel, Linear systems, Vol.~8, Springer, 1997.

\bibitem{generic}
J.~M. Dion, C.~Commault, J.~Van~DerWoude, Generic properties and control of
  linear structured systems: {a} survey, Automatica 39 (2003) 1125--1144.

\bibitem{kaplan1966introduction}
W.~Kaplan, Introduction to {Analytic Functions}, Addison-Wesley Publishing
  Company, 1966.

\bibitem{R.A.1991Topics}
R.~A. Horn, C.~R. Johnson, Topics in Matrix Analysis, Cambridge University
  Press, 1991.

\bibitem{gao2014target}
J.~Gao, Y.-Y. Liu, R.~M. D'souza, A.-L. Barab{\'a}si, Target control of complex
  networks, Nature communications 5~(1) (2014) 1--8.

\bibitem{czeizler2018structural}
E.~Czeizler, K.-C. Wu, C.~Gratie, K.~Kanhaiya, I.~Petre, Structural target
  controllability of linear networks, IEEE/ACM Transactions on Computational
  Biology and Bioinformatics 15~(4) (2018) 1217--1228.

\bibitem{murota1990note}
K.~Murota, S.~Poljak, Note on a graph-theoretic criterion for structural output
  controllability, IEEE Transactions on Automatic Control 35~(8) (1990)
  939--942.

\bibitem{li2020structural}
J.~Li, X.~Chen, S.~Pequito, G.~J. Pappas, V.~M. Preciado, On the structural
  target controllability of undirected networks, IEEE Transactions on Automatic
  Control 66~(10) (2020) 4836--4843.

\bibitem{Murota_Book}
K.~Murota, Matrices and Matroids for Systems Analysis, Springer Science
  Business Media, 2009.

\bibitem{poljak1989maximum}
S.~Poljak, Maximum rank of powers of a matrix of a given pattern, Proceedings
  of the American Mathematical Society 106~(4) (1989) 1137--1144.

\bibitem{lawler2001combinatorial}
E.~L. Lawler, Combinatorial Optimization: Networks and Matroids, Courier
  Corporation, 2001.

\bibitem{hosoe1980determination}
S.~Hosoe, Determination of generic dimensions of controllable subspaces and its
  application, IEEE Transactions on Automatic Control 25~(6) (1980) 1192--1196.

\bibitem{nemhauser1978analysis}
G.~L. Nemhauser, L.~A. Wolsey, M.~L. Fisher, An analysis of approximations for
  maximizing submodular set functions¡ªi, Mathematical programming 14 (1978)
  265--294.

\bibitem{P.Fa2014Controllability}
F.~Pasqualetti, S.~Zampieri, F.~Bullo, Controllability metrics, limitations and
  algorithms for complex networks, IEEE Transactions on Control of Network
  Systems 1~(1) (2014) 40--52.

\bibitem{ji2008controllability}
Z.~Ji, L.~Wang, X.~Guo, On controllability of switched linear systems, IEEE
  Transactions on Automatic Control 53~(3) (2008) 796--801.

\end{thebibliography}
		}
	\end{document}